\documentclass[longauth]{aa} 
\usepackage[varg]{txfonts} 
\usepackage{graphicx}
\usepackage{natbib}
\bibpunct{(}{)}{;}{a}{}{,} 
\usepackage{multirow}
\usepackage{amssymb} 
\usepackage{tikz} 
\usepackage{hyperref}
\hypersetup{
colorlinks,
citecolor=blue,
}
\usepackage{xcolor}

\DeclareMathOperator*{\median}{median}

\defcitealias{HC21}{HC21} 

\begin{document}

\title{J-NEP: 60-band photometry and photometric redshifts for the {\it James Webb Space Telescope} North Ecliptic Pole Time-Domain Field}

\author{A.~Hern\'an-Caballero\inst{\ref{CEFCA2} \thanks{email: ahernan@cefca.es}}
\and C.~N.~A.~Willmer\inst{\ref{Arizona}}
\and J.~Varela\inst{\ref{CEFCA}}
\and C.~L\'opez-Sanjuan\inst{\ref{CEFCA2}}
\and A.~Mar\'in-Franch\inst{\ref{CEFCA2}}
\and H.~V\'azquez Rami\'o\inst{\ref{CEFCA2}}
\and T.~Civera\inst{\ref{CEFCA}}
\and A.~Ederoclite\inst{\ref{CEFCA2}}
\and D.~Muniesa\inst{\ref{CEFCA}}
\and J.~Cenarro\inst{\ref{CEFCA2}}
\and S.~Bonoli\inst{\ref{CEFCA},\ref{DIPC},\ref{Iker}}
\and R.~Dupke\inst{\ref{ON},\ref{UMich},\ref{UAlab}}
\and J.~Lim\inst{\ref{HongKong}}
\and J.~Chaves-Montero\inst{\ref{DIPC}}
\and J.~Laur\inst{\ref{Tartu}}
\and C.~Hern\'andez-Monteagudo\inst{\ref{ULL},\ref{IAC}}
\and J.~A.~Fern\'andez-Ontiveros\inst{\ref{CEFCA2}}
\and A.~Fern\'andez-Soto\inst{\ref{IFCA},\ref{IFCA-UV}}
\and L.~A.~D\'iaz-Garc\'ia\inst{\ref{IAA}}
\and R.~M.~Gonz\'alez Delgado\inst{\ref{IAA}}
\and C.~Queiroz\inst{\ref{IF/USP},\ref{UFRGS}}
\and J.~M.~V\'ilchez\inst{\ref{IAA}}
\and R.~Abramo\inst{\ref{IF/USP}}
\and J.~Alcaniz\inst{\ref{ON}}
\and N.~Ben\'itez\inst{\ref{IAA}}
\and S.~Carneiro\inst{\ref{IF/UFB}}
\and D.~Crist\'obal-Hornillos\inst{\ref{CEFCA}}
\and C.~Mendes de Oliveira\inst{\ref{SaoPaulo}}
\and M.~Moles\inst{\ref{CEFCA},\ref{IAA}}
\and L.~Sodr\'e Jr.\inst{\ref{SaoPaulo}}
\and K.~Taylor\inst{\ref{Inst4}}
}
\institute{Centro de Estudios de F\'isica del Cosmos de Arag\'on (CEFCA), Plaza San Juan, 1, E-44001 Teruel, Spain\label{CEFCA2}
\and Steward Observatory, University of Arizona, 933 North Cherry Avenue, Tucson, AZ 85721, USA\label{Arizona}
\and Centro de Estudios de F\'isica del Cosmos de Arag\'on (CEFCA), Unidad Asociada al CSIC, Plaza San Juan, 1, E-44001 Teruel, Spain\label{CEFCA}
\and Donostia International Physics Centre, Paseo Manuel de Lardizabal 4, E-20018 Donostia-San Sebastian, Spain\label{DIPC}
\and Ikerbasque, Basque Foundation for Science, E-48013 Bilbao, Spain\label{Iker}
\and Observat\'orio Nacional, Minist\'erio da Ci\^encia, Tecnologia, Inova\c{c}\~ao e Comunica\c{c}\~oes, Rua General Jos\'e Cristino, 77, S\~ao Crist\'ov\~ao, 20921-400, Rio de Janeiro, Brazil\label{ON}
\and Department of Astronomy, University of Michigan, 311 West Hall, 1085 South University Ave., Ann Arbor, USA\label{UMich}
\and Department of Physics and Astronomy, University of Alabama, Box 870324, Tuscaloosa, AL, USA\label{UAlab}
\and Department of Physics, University of Hong Kong, 852-2219 4924, HK\label{HongKong}
\and Tartu Observatory, University of Tartu, Observatooriumi 1, 61602 T\~oravere, Estonia\label{Tartu}
\and Departamento de Astrof\'isica, Universidad de La Laguna, E-38206, La Laguna, Tenerife, Spain\label{ULL}
\and Instituto de Astrof\'isica de Canarias, E-38200 La Laguna, Tenerife, Spain\label{IAC} 
\and Instituto de F\'{\i}sica de Cantabria (Consejo Superior de Investigaciones Cient\'{\i}ficas - Universidad de Cantabria). Avda. de los Castros, E-39005, Santander, Spain\label{IFCA}
\and Unidad Asociada ``Grupo de astrof\'{\i}sica extragal\'actica y cosmolog\'{\i}a'' (IFCA - Universitat de Val\`encia). Parc Cient\'{\i}fic UV, E-46980, Paterna, Spain\label{IFCA-UV} 
\and Instituto de Astrof\'{\i}sica de Andaluc\'{\i}a (CSIC), P.O.~Box 3004, E-18080 Granada, Spain\label{IAA}
\and Instituto de F\'isica, Universidade de S\~ao Paulo, Rua do Mat\~ao 1371, CEP 05508-090, S\~ao Paulo, Brazil\label{IF/USP}
\and Departamento de Astronomia, Instituto de F\'isica, Universidade Federal do Rio Grande do Sul (UFRGS), Av. Bento Gon\c{c}alves 9500,
Porto Alegre, R.S, Brazil\label{UFRGS}
\and Instituto de F\'isica, Universidade Federal da Bahia, 40210-340, Salvador, BA, Brazil\label{IF/UFB}
\and Departamento de Astronomia, Instituto de Astronomia, Geof\'isica e Ci\^encias Atmosf\'ericas da USP, Cidade Universit\'aria, 05508-900, S\~ao Paulo, SP, Brazil\label{SaoPaulo}
\and Instruments4, 4121 Pembury Place, La Ca\~nada-Flintridge, CA, 91011, USA\label{Inst4}
}

\date{Accepted ........ Received ........;}

\abstract {
The Javalambre-Physics of the Accelerating Universe Astrophysical Survey (J-PAS) will observe $\sim$1/3 of the northern sky with a set of 56 narrow-band filters using the dedicated 2.55 m Javalambre Survey Telescope (JST) at the Javalambre Astrophysical Observatory. Prior to the installation of the main camera, in order to demonstrate the scientific potential of J-PAS, two small surveys were performed with the single-CCD Pathfinder camera: miniJPAS ($\sim$1 deg$^2$ along the Extended Groth Strip), and J-NEP ($\sim$0.3 deg$^2$ around the JWST North Ecliptic Pole Time Domain Field), including all 56 J-PAS filters as well as $u$, $g$, $r$, and $i$. J-NEP is $\sim$0.5--1.0 magnitudes deeper than miniJPAS, providing photometry for 24,618 $r$-band detected sources and photometric redshifts (photo-$z$) for the 6,662 sources with $r$$<$23. 

In this paper we describe the photometry and photo-$z$ of J-NEP and demonstrate a new method for the removal of systematic offsets in the photometry based on the median colours of galaxies, dubbed ``galaxy locus recalibration''. This method does not require spectroscopic observations except in a few reference pointings and, unlike previous methods, is directly applicable to the whole J-PAS survey.

We use a spectroscopic sample of 787 galaxies to test the photo-$z$ performance for J-NEP and in comparison to miniJPAS.
We find that the deeper J-NEP observations result in a factor $\sim$1.5--2 decrease in $\sigma_{\rm{NMAD}}$ (a robust estimate of the standard deviation of the photo-$z$ error) and $\eta$ (the outlier rate) relative to miniJPAS for $r$$>$21.5 sources, but no improvement in brighter ones probably due to systematics.
We find the same relation between $\sigma_{\rm{NMAD}}$ and $odds$ in J-NEP and miniJPAS, which suggests that we will be able to predict the $\sigma_{\rm{NMAD}}$ of any set of J-PAS sources from their $odds$ distribution alone, with no need for additional spectroscopy to calibrate the relation.
We explore the causes for photo-$z$ outliers and find that colour-space degeneracy at low S/N, photometry artefacts, source blending, and exotic spectra are the most important factors.
} 


\titlerunning{Photometry and photo-$z$ of J-NEP}
\authorrunning{A. Hern\'an-Caballero et al.}

\maketitle

\section{Introduction} 

The Javalambre-Physics of the Accelerating Universe Astrophysical Survey \citep[J-PAS;][]{Benitez09,Benitez14} will observe thousands of square degrees in the Northern hemisphere with a unique set of 56 optical filters that provides, for each $ 0 \farcs 48 \times 0 \farcs 48$ pixel on the sky, an $R$$\sim$60  photo-spectrum (hereafter j-spectrum) covering the 3800--9100 \AA{} range. 
This will enable achieving the extremely accurate photometric redshifts (photo-$z$),
needed to measure the cosmological baryonic acoustic oscillation (BAO) signal across a wide range of cosmic epochs, and to characterise millions of stars, galaxies and quasars to a level of detail previously restricted to spectroscopic studies.

The instrument that makes J-PAS possible, JPCam, has 14 CCDs (each containing 9200$\times$9200 pixels) covering an area of $\sim$4.2 deg$^2$ in a single pointing. JPCam is currently being commissioned on the 2.5 m JST/T250 telescope at the Observatorio Astrof\'isico de Javalambre \citep[OAJ;][]{Cenarro14}. 

Prior to the installation of JPCam, a single-CCD camera (Pathfinder) was used on the JST/T250 telescope to start its scientific operation and to provide observations with the same filter set of J-PAS in advance of the actual survey.
Most of the observing time with Pathfinder was devoted to performing a $\sim$1 deg$^2$ survey of the Extended Groth Strip with the 56 filters of J-PAS along with $u$, $g$, $r$ and $i$, reaching depths comparable to those expected for J-PAS \citep{Bonoli21}. This survey, dubbed miniJPAS, has been used to test the performance of the telescope and the Pathfinder camera \citep{Bonoli21, Xiao21}, 
the algorithms for detection of emission lines \citep{Martinez-Solaeche21,Iglesias-Paramo22}, star/galaxy/quasar classification \citep{Baqui21,Queiroz22}, and photo-$z$ calculation \citep{HC21,Laur22}, as well as to obtain the first scientific results on the properties of J-PAS galaxies \citep{Gonzalez-Delgado21,Gonzalez-Delgado22,Martinez-Solaeche22,Rodriguez-Martin22} and quasars \citep{Chaves-Montero22}.

A second survey performed with Pathfinder is the Javalambre North Ecliptic Pole Survey (J-NEP). Like miniJPAS, J-NEP uses the 56 filters of J-PAS plus $u$, $g$, $r$, and $i$. It consists of a single Pathfinder pointing ($\sim$0.3 deg$^2$) but with significantly longer total exposure times, reaching between $\sim$0.5 and $\sim$1.0 magnitudes deeper than miniJPAS (see Sect. \ref{sec:data}).

J-NEP covers entirely the {\it James Webb Space Telescope} (JWST) North Ecliptic Pole Time-Domain Field \citep[JWST-TDF;][]{Jansen18}. The JWST-TDF is a new extragalactic field of interest selected for its low Galactic extinction, absence of bright stars, and location within the Northern continuous viewing zone of JWST. The guaranteed time program JWST-GTO-2738 (PI: Windhorst) has been awarded $\sim$47 hours to imaging the JWST-TDF in 8 filters with JWST/NIRCam and JWST/NIRISS slit-less spectroscopy in parallel, drawing a 4-spoke pattern (Fig. \ref{fig:footprint}). 
The multi-wavelength coverage ranges from hard X-rays to radio wavelengths as tabulated by R. Jansen\footnote{\url{http://lambda.la.asu.edu/jwst/neptdf/}}. It includes:
broad-band imaging in the optical and near-infrared with the \textit{Hubble Space Telescope} (PI: Jansen), \textit{Subaru} (PIs: Hasinger \& Hu), and the \textit{Gran Telescopio Canarias} (PI: Dhillon); X-rays with \textit{NuSTAR} (PI: Civano) and \textit{Chandra} (PI: Maksym); submillimetre observations with the \textit{James Clerk Maxwell Telescope} (PIs: Smail \& Im) and \textit{IRAM 30m} (PI: Cohen), radio observations with the \textit{Very Large Array} (PIs: Windhorst \& Cotton), \textit{Very Long Baseline Array} (PI: Brisken), and the \textit{Low-Frequency Array} (PI: Van Weeren), among others.

A significant effort to obtain spectroscopy in the JWST-TDF is also ongoing. Willmer et al. (in prep.) obtained 1553 \textit{MMT}/Binospec spectra covering the $\sim$3900--9350 \AA{} wavelength range, 1013 of them yielding high confidence spectroscopic redshifts. 
However, for the majority of sources within the J-NEP footprint, photo-$z$ will be the only redshift estimates available in the foreseeable future. The J-NEP j-spectra will thus play an important role in the characterisation of many of the sources detected in the multi-wavelength surveys. 

Unlike miniJPAS and J-NEP, most of the area that will be covered by J-PAS lacks any substantial spectroscopic coverage at the depth required to validate the photo-$z$ ($r$$\sim$23). As a consequence, the strategies for prediction of the photo-$z$ performance and for removal of systematic offsets in the photometry \citep[see][hereafter \citetalias{HC21}]{HC21} need to be made independent from the availability of spectroscopy at a given pointing.

In this paper, we present the photometry and photo-$z$ of the J-NEP survey.
In preparation for J-PAS, and in contrast to miniJPAS, the photo-$z$ of J-NEP have been obtained independently of any spectroscopic information in all the steps of the process (the team responsible for the data reduction, photometry, and photo-$z$ calculation only had access to the spectroscopic data after delivery of the final photo-$z$ catalogue). This eliminates the possibility of overestimating the actual photo-$z$ accuracy that in miniJPAS could result from the overlap among the spectroscopic samples used for template selection, photometric recalibration, and photo-$z$ validation.

Section \ref{sec:data} describes the J-NEP observations, the data reduction and calibration, as well as the spectroscopic redshifts used to validate the J-NEP photo-$z$. Section \ref{sec:star-gal-class} discusses the star/galaxy classification of J-NEP. In Sect. \ref{sec:photo-z-main}, we present the methodology for photo-$z$ calculation and compare the distribution of photo-$z$ in J-NEP and miniJPAS. In Sect. \ref{sec:accuracy}, we evaluate the accuracy of photo-$z$ in J-NEP using the spectroscopic subsample, compare them to miniJPAS, and extrapolate to the entire J-NEP sample. Sect. \ref{sec:outliers} discuses the causes for photo-$z$ outliers and Sect. \ref{sec:summary} summarises our conclusions. All magnitudes are presented in the AB system. 

\section{Observations and data reduction}\label{sec:data}

\subsection{J-NEP observations}

\begin{figure} 
\begin{center}
\includegraphics[width=8.4cm]{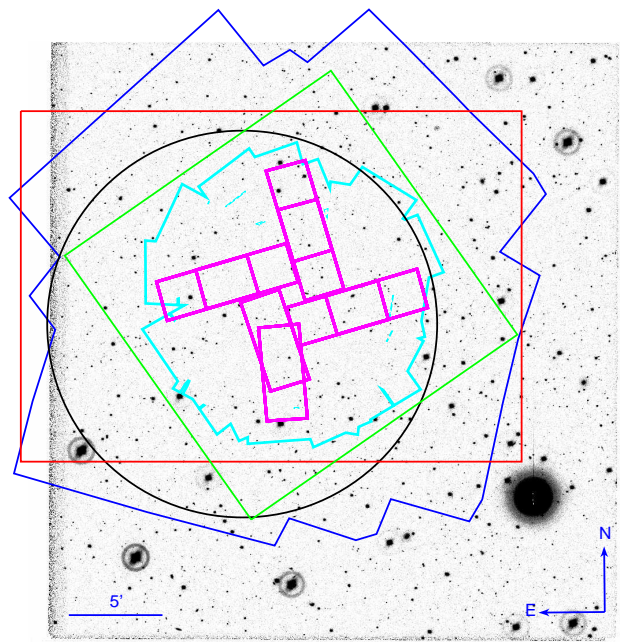}
\end{center}
\caption[]{Co-added $r$-band image from J-NEP.  
The 4-spoke pattern in magenta indicates the regions with NIRCam+NIRISS coverage from the JWST-GTO-2738 program \citep[see][for details]{Jansen18}, including fields observed in 2022 (southern and eastern spokes) and planned for 2023 (northern and western).
Other lines mark the footprint of ancillary observations with HST/ACS+WFC3 (cyan polygon), NuSTAR (dark blue polygon), Chandra (green square), VLA (black circle), and the area covered by spectroscopic observations with MMT/Binospec (red rectangle).\label{fig:footprint}}
\end{figure}

The J-NEP survey is centred at (RA, Dec) = (17$^h$22$^m$26$^s$,+65$^d$46$^m$48$^s$), covering the entirety of the JWST-TDF footprint as well as most of the ancillary observations (see Figure \ref{fig:footprint}). 
Observations of the J-NEP were carried out between 23 June 2018 and
23 July 2019. Observations in broad bands were executed every time a new
set of narrow band filters were swapped, resulting in dozens of
exposures in the $g$, $r$, and $i$ bands.
The coadded images for the $g$ and $i$ bands include all the exposures taken, totalling 151 images (4530 s total exposure time) and 96 images (2880 s), respectively.
For the coadded image in the $r$ band (which is the reference for source extraction and forced photometry, see Sect. \ref{sec:photometry}) only the 99 images with point spread functions (PSF) full width at half-maximum (FWHM) $<$ 1.0\arcsec were combined, resulting in a coadded image of 2970 s exposure time. 
For the narrow/medium bands, the number of exposures combined ranges between 4 and 52 (total exposure times between 480 s and 6240 s). In contrast
to miniJPAS, the J-NEP observations did not try to reproduce the nominal depth of J-PAS observations and the field was observed whenever possible. 

\begin{figure} 
\begin{center}
\includegraphics[width=8.4cm]{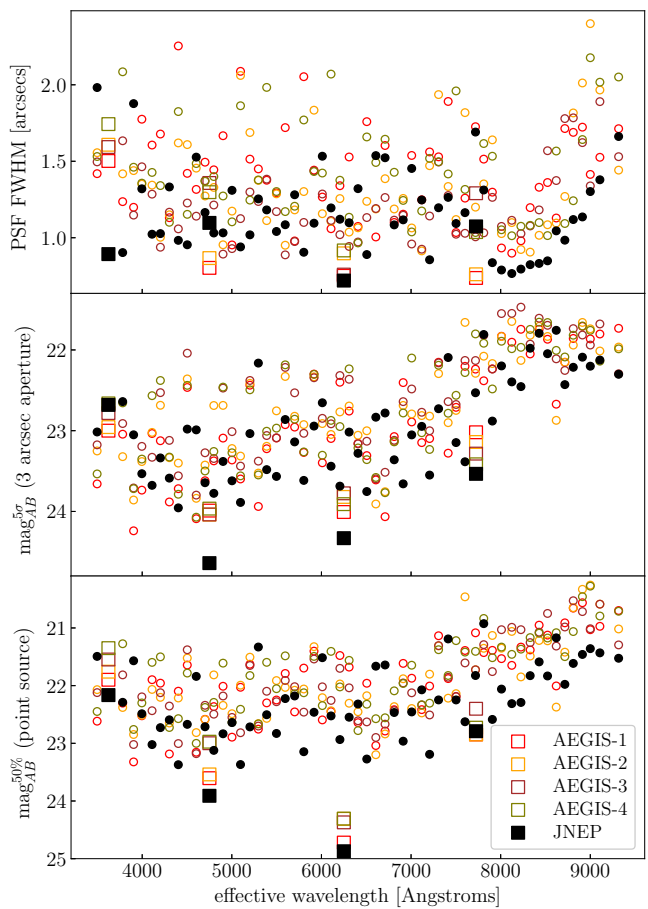}
\end{center}
\caption[]{Comparison of image PSF FWHM (top), depth (middle), and magnitude at 50\% completeness for point sources (bottom). Solid black symbols indicate values for J-NEP, while open coloured symbols represent each of the four pointings of miniJPAS. Big squares correspond to the broad-bands $u$, $g$, $r$, $i$, while small circles correspond to the 56 bands of J-PAS.\label{fig:FWHM-depth}}
\end{figure} 

Figure \ref{fig:FWHM-depth} shows the PSF FWHM of the co-added J-NEP images in each of the filters, the 5-$\sigma$ depths computed from the noise in 3\arcsec circular apertures (mag$^{5\sigma}_{AB}$), and the 50\% completeness limit for point sources that is estimated by injecting simulated sources in the images (mag$^{50\%}_{AB}$). Values for the four individual miniJPAS pointings (labels AEGIS-1 to AEGIS-4) are also shown for comparison. 
The PSF FWHM of J-NEP ranges from 0.72\arcsec ($r$ band) to 1.98\arcsec ($u$ band). The FWHM is smaller than the average for miniJPAS in most of the bands, and especially in the reddest ones, which for miniJPAS were observed at low elevation \citep{Bonoli21}. J-NEP images are also deeper due to longer total exposure times (median $\Delta$mag$^{5\sigma}_{AB}$ = 0.37). This results in 
50\% completeness limits that are $\sim$0.5 mag fainter, on average, compared to miniJPAS (median $\Delta$mag$^{50\%}_{AB}$ = 0.56). 

\subsection{Image processing and aperture photometry}\label{sec:photometry}

The reduction of J-NEP images, source detection, and aperture photometry used the same procedure as for the miniJPAS. A detailed description of the full process is presented in \citet{Bonoli21}.

Data reduction for the individual images includes the standard bias and over-scan subtraction, trimming, flat fielding, and illumination correction. In the red bands, a fringing correction is also applied. Observations with the Pathfinder camera also require additional corrections for vignetting and background patterns \citep[see][for details]{Bonoli21}.
Astrometric calibration was performed with {\sc Scamp} \citep{Bertin06} using the Gaia DR2 catalogue as reference. Coaddition of the individual images was performed with {\sc Swarp} \citep{Bertin02}, with all the images resampled to the fiducial pixel scale of the camera (0.23\arcsec). 

Source detection and extraction on the co-added images were performed with {\sc SExtractor} \citep{Bertin96}. Aperture photometry was obtained in both dual-mode (the reference band, $r$, defines the aperture in which photometry is extracted for all the bands) and single-mode (source detection and extraction performed independently in each band) for several aperture types and sizes. 
The dual-mode catalogue contains photometry in all the filters and apertures for the 24,618 sources detected in the $r$-band image.

Following \citetalias{HC21}, we use magnitudes in the AUTO aperture (2 Kron radii) as a proxy for total magnitudes. However, colour indices and photo-$z$ are computed using PSFCOR magnitudes, which represent the magnitude that would be measured in a ``restricted AUTO'' aperture (1 Kron radius) if the PSF of the image were equal to the PSF of the reference band \citep[see][for a detailed description of the method]{Coe06,Molino19}. 

Photometry flags from {\sc SExtractor} (which indicate different issues that may degrade or invalidate the photometry) are defined separately for every source in each band. The fraction of sources affected by flags is higher than in miniJPAS (30\% vs 24\% for $r$$<$23 sources), probably due to the higher density of foreground stars (see Sect. \ref{sec:star-gal-class}).

Photometric calibration, including corrections for atmospheric extinction was performed using an adaptation of the method presented in \citet{Lopez-Sanjuan19a}. We applied the following steps:

1) Selection of a high-quality sample of stars for calibration. We selected all J-NEP sources with a $S/N > 10$ in all photometric bands and for which a $S/N > 3$ parallax from \textit{Gaia} DR2 is available. Using \textit{Gaia} photometry, we constructed a diagram of the absolute $G$ magnitude versus $(G_{\rm BP} - G_{\rm RP})$ colour and de-reddened for dust using the 3D extinction maps from \citet{Green18}. We then selected those sources belonging to the main sequence, yielding $175$ calibration stars.

2) Calibration of the images in the $g$, $r$, and $i$ bands using the Pan-STARRS photometry as reference. We compared our $6$ arcsec diameter aperture magnitudes corrected to total magnitudes with the PSF magnitudes from Pan-STARRS. The aperture corrections were computed from the light growth curves of unsaturated, bright stars in each tile and are stored in the table \textit{jnep.TileImage} of the J-NEP database. This step provides the zero points of the images in the $g$, $r$, and $i$ bands. 

3) Calibration of the narrow bands with the stellar locus. For each narrow band, we computed the dust de-reddened $(\mathcal{X}_{\rm ins} - r)_0$ vs. $(g - i)_0$ colour-colour diagram of the calibration stars, where $\mathcal{X}_{\rm ins}$ is the instrumental magnitude of the selected narrow band and the extinction coefficients $k_{\mathcal{X}}$ were calculated for the extinction law of \citet{Schlafly16} using the prescription in \citet{Whitten19}. We computed the offsets needed for making the J-NEP stellar locus consistent with a reference stellar locus estimated using the miniJPAS photometry. The initial offset was adjusted to account for the variation in the stellar locus position due to the different average metallicity of the calibration stars between the miniJPAS and the J-NEP pointings using the trends found by \citet{Lopez-Sanjuan21} in J-PLUS. This process yields the zero points of the narrow band filters.
Following the results in \citet{Bonoli21}, we believe that the photometric calibration has an absolute error of at most 0.04 mag.

\subsection{Recalibration with the Galaxy Locus}\label{sec:recalibration}

\begin{figure} 
\begin{center}
\includegraphics[width=8.4cm]{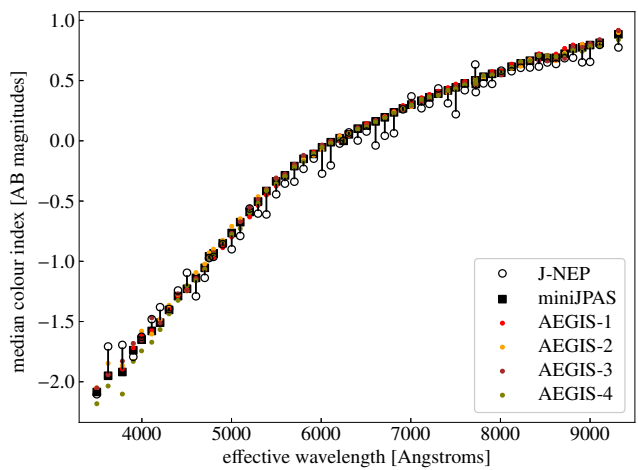}
\end{center}
\caption[]{Galaxy Locus for $r$$<$21 galaxies, defined as the median magnitude difference relative to the $r$ band for each of the J-PAS bands. Coloured dots represent the Galaxy Locus obtained from the recalibrated PSFCOR photometry on individual miniJPAS pointings, while the solid squares indicate the average value in the four pointings. Open circles mark the Galaxy Locus obtained from PSFCOR photometry in J-NEP before recalibration. The segments connecting them to the squares represent the corrections required to remove systematic offsets in the PSFCOR photometry of J-NEP.\label{fig:galaxy-locus}}
\end{figure} 

We compute colour indices and photo-$z$ using the dual-mode photometry with PSF-corrected apertures \citepalias[PSFCOR; see Sect. 2.2 in][for details]{HC21}. Since PSFCOR fluxes are measured in a relatively small aperture (1 Kron radius), we scale the fluxes in all bands by a factor defined as the ratio between AUTO and PSFCOR fluxes in the $r$ band. This makes PSFCOR magnitudes closer to total magnitudes if we dismiss the effects of radial colour gradients. The upscaling of PSFCOR fluxes is required in order to apply the correct redshift prior (which depends on the total magnitude of the source, see Sect. \ref{sec:photo-z}). 

The PSFCOR photometry does not completely remove the effect of PSF variations in the co-added images in different bands. This results in systematic offsets in the colours of the sources. Since accurate colours are essential for high precision photo-$z$, \citetalias{HC21} proposed a method for the recalibration of PSFCOR photometry using stellar population synthesis (SPS) model fits to the j-spectra of individual galaxies. 
In this procedure, the observed magnitudes in the J-PAS bands (corrected for atmospheric and Galactic extinction) of galaxies covering a wide redshift range are compared to synthetic photometry obtained by convolving the best-fitting SPS model for each galaxy with the transmission profiles of the filters. Systematic offsets in the photometry are computed as the median magnitude difference between the observed and synthetic photometry in each band. In an iterative process, the observed photometry is corrected by subtracting these offsets, then fitting new SPS models to the corrected photometry from which synthetic photometry is obtained and additional offsets are computed. The process converges after 3 to 4 iterations (see Sect. 3 in \citetalias{HC21} for details).

Such iterative SPS model fitting recalibration (hereafter, ISMFR) produces excellent results, with colours accurate to within $\sim$0.02 mag. However, it depends critically on the availability of spectroscopic redshifts (that are needed to fit the SPS models) for a large number of galaxies in every pointing.
Since the spectroscopic coverage in most J-PAS pointings will be insufficient for a recalibration à la miniJPAS, we developed an alternative recalibration strategy that we call galaxy locus recalibration (GLR).
The GLR method relies on ISMFR to obtain accurate colours for the sources in some reference pointings with spectroscopy (in our case, the 4 miniJPAS pointings). 

In analogy with the stellar locus, we define the galaxy locus as a set of colour indices \{$\tilde{C}_j$\}, where each element $\tilde{C}_j$ is the median colour between the $j$-th band and the $r$ band

\begin{equation}
\tilde{C}_j = \median_{i \in R} \big{\{} m_i(r) - m_i(j)\big{\}}
\end{equation}

\noindent The set $R$ indexes all the galaxies brighter than a given limit, $m_\textrm{cut}$. For the determination of the galaxy locus, we consider as galaxies all the sources that have {\sc SExtractor} class\_star $<$ 0.1. 
We set $m_\textrm{cut}$ = 21. There are 3067 sources in miniJPAS meeting the criteria $r$$<$21 and class\_star $<$ 0.1. A fainter $m_\textrm{cut}$ increases the number of galaxies available to compute the $\tilde{C}_j$ indices, but the average uncertainty in the colours of individual galaxies also increases, particularly in the bluest bands. We find that $m_\textrm{cut}$$\sim$21 minimises the statistical uncertainty of the $\tilde{C}_j$.

To recalibrate J-NEP with the galaxy locus, we assume that the intrinsic colours of galaxies (that is, the colours obtained after correcting for Galactic extinction and any systematics in the photometry) are independent from the sky coordinates. 
Therefore, we can estimate the systematic offsets in the J-NEP photometry as $\Delta m(j)$ = $\tilde{C}_j$ - $\tilde{C}_j^{obs}$, 
where $\tilde{C}_j^{obs}$ is the ``apparent'' galaxy locus obtained from the observed J-NEP photometry (after correcting for Galactic extinction).

Figure \ref{fig:galaxy-locus} compares the values of the \{$\tilde{C}_j$\} and  \{$\tilde{C}_j^{obs}$\}. The difference can be as large as $\vert\Delta m(j)\vert$$\sim$0.3 mag, which is consistent with the similarly large recalibration offsets obtained for miniJPAS with ISMFR \citepalias{HC21}.
The GLR method is less accurate than ISMFR because it depends on the latter to determine the galaxy locus and is affected by other sources of uncertainty such as the statistical error in the median colours obtained from a limited sample. Cosmic variance could also introduce a systematic error in the GLR method if the galaxy population or its redshift distribution in the pointing to be recalibrated are significantly different from the reference. 
However, in Appendix \ref{sec:appendix} we evaluate each contribution to the total uncertainty of the GLR method, finding a negligible impact from cosmic variance. We predict a total systematic error in J-NEP colours of $\sim$0.04 mag. This is a factor of $\sim$2 worse than ISMFR achieves, but a factor of $\sim$4--5 improvement with respect to no recalibration. To compute photo-$z$, we add this systematic error in quadrature to the photometric uncertainties.

A systematic error of $\sim$0.04 mag in the colours implies that the photometric error dominates over systematics for most sources. The impact of the recalibration uncertainty on the photo-$z$ is therefore expected to be small, except possibly for bright sources with high S/N photometry but weak spectral features, where small offsets in the photometry could mimic the effect of emission or absorption lines. In Sect. \ref{sec:validation}, we show what is the actual impact of this uncertainty in the photo-$z$ of bright sources.

\subsection{Spectroscopic observations}

The spectroscopic observations were obtained using Binospec
\citep{Fabricant19} at the MMT observatory. They used a
270~lines/mm grating  which allows covering from approximately
4000~\AA{} to  9000~\AA{} with a typical dispersion of 1.30~\AA/pixel and
resolution of 1340 at the central wavelength. The sample
of galaxies was selected from a combination of a preliminary catalogue
coming from the HEROES Subaru HSC imaging (Hasinger, private communication) with
a catalogue derived from MMT/MMIRS near-infrared imaging (Willmer et
al. in preparation), limited at \textsl{r} $\sim$ 23.
The data were reduced using a specially designed pipeline that
produces wavelength and flux calibrated 1-dimensional spectra
\citep{Kansky19}. Redshifts were measured using custom-written
code using a combination of real-space cross-correlation and
emission-line fits. All spectra were visually inspected and
a redshift quality flag was assigned using the same criteria adopted by the DEEP2 survey \citep{Newman13}, where a redshift quality of 4 (3) represents a $>$95\% ($>$90\%) confidence in the spectroscopic redshift. 
The outputs from each of the observed fields were then
merged into a single list using the positions of the near-infrared
imaging, which are tied to the \textit{Gaia} DR2 reference frame.

The spectroscopic redshift catalogue contains 1553 sources.
We associate spectroscopic redshifts to J-NEP sources by looking for $r$$<$23 counterparts within 1\arcsec of the spectroscopic position. We find matches for 955 sources (the remainder are either fainter than $r$=23 or outside the J-NEP footprint). The median distance of the associations is $\sim$0.1\arcsec, with no significant systematic offset in the coordinates, and 95\% of the associations are within 0.4\arcsec.
For the purpose of validation of the photo-$z$, only robust spectroscopic redshifts (ZQUALITY$\ge$3) are considered. This decreases the number of useful spectroscopic redshifts to 787. 
The spectroscopic catalogue contains no spectral classification. We consider the 58 sources with $z_{\rm{spec}}$$<$0.001 to be likely stars and excluded them from the analysis of photo-$z$ performance. All but six of these sources have SExtractor CLASS\_STAR $>$ 0.9 and/or point-source probabilities $P_{star}$ $>$ 0.9 (see Sect. \ref{sec:star-gal-class}). There are also 18 point sources (CLASS\_STAR $>$ 0.9 and/or $P_{star}$ $>$ 0.9) with $z_{\rm{spec}}$$>$0.001. We inspected their spectra and found six of them to be quasars with clear broad emission lines (3 of them are detected by NuSTAR and all 6 have WISE colours consistent with a quasar). We also excluded them from the analysis. The remaining 12 point sources with $z_{\rm{spec}}$$>$0.001 have spectra consistent with compact galaxies and remain in the validation sample. 

We emphasise that the spectroscopic information was not made available to the team processing the J-NEP data until after the star/galaxy classification (see Sect. \ref{sec:star-gal-class}) and photo-$z$ calculation (Sect. \ref{sec:photo-z}) were finished. We purposefully avoid updating the star/galaxy classification and photo-$z$ of J-NEP sources in light of the spectroscopic information. Instead, we use the spectroscopic redshifts only for validation of the photo-$z$ performance.

\section{Star and galaxy classification\label{sec:star-gal-class}}

\begin{figure} 
\begin{center}
\includegraphics[width=8.4cm]{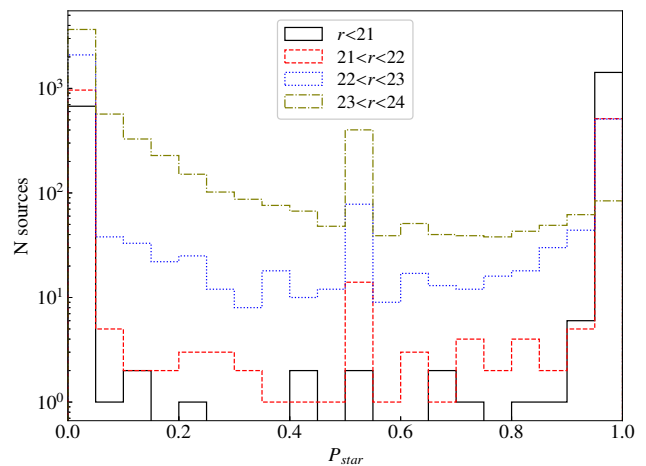}
\end{center}
\caption[]{Distribution of the probability of being a compact source ($P_{star}$) computed with the SGLC method for J-NEP sources in several magnitude intervals. The spike at $P_{star}$ = 0.5 is due to sources with insufficient S/N for a classification in the detection band, which are assigned $P_{star}$ = 0.5 by default.\label{fig:pstar-distrib}}
\end{figure}

In addition to the morphological classification from {\sc SExtractor} (the CLASS\_STAR parameter), for every source in J-NEP we assign a probability $P_{star}$ of being a point source using the Stellar-Galaxy Locus Classification (SGLC) described in \citet{Lopez-Sanjuan19b}. This method uses Bayesian inference to estimate the probability of a given source being a compact object (star or quasar) from the value and uncertainty of its concentration parameter, $c$, defined as the difference between the 1.5\arcsec and 3\arcsec diameter aperture magnitudes of the source. The star probability was computed in the $g$, $r$, and $i$ bands and the values were combined assuming independent measurements. In addition, a prior on the r-band magnitude (and for bright sources, also parallaxes from \textit{Gaia} DR2) was applied. The resulting distribution of $P_{star}$ is strongly bimodal, with most sources unambiguously classified as either compact or extended (see Fig. \ref{fig:pstar-distrib}).

\begin{figure} 
\begin{center}
\includegraphics[width=8.4cm]{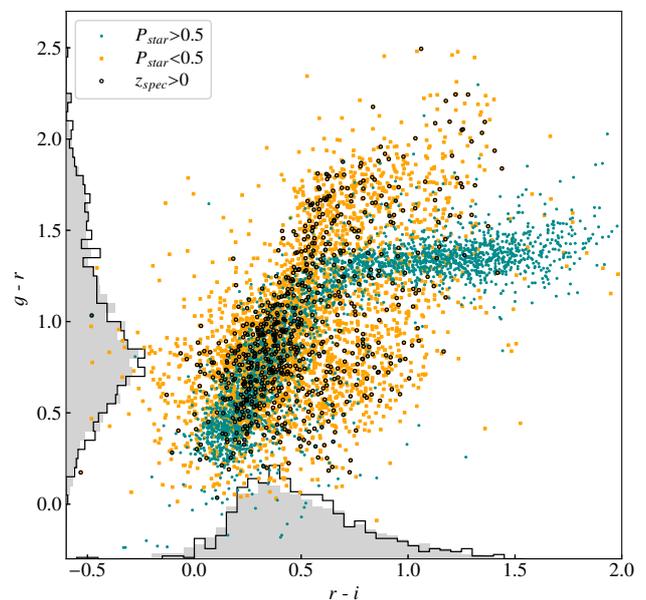}
\end{center}
\caption[]{Colour-colour plot including all J-NEP sources with $r$$<$23 and S/N$>$2 in the $g$, $r$ and $i$ bands. Cyan dots and orange squares represent sources morphologically classified as stars and galaxies, respectively. Black circles indicate the spectroscopically confirmed extragalactic sources (including some quasars). The solid grey and step black line histograms represent the distribution of the colours for morphologically and spectroscopically classified galaxies, respectively.\label{fig:gri-star-galaxy}}
\end{figure} 

It is important to emphasise that no colour information is used in the classification. As discussed in \citetalias{HC21}, this implies that the redshift probability distribution function $P$($z$) derived from the j-spectrum is independent of $P_{star}$. 

Figure \ref{fig:gri-star-galaxy} compares the $g$-$r$ and $r$-$i$ colours for individual sources with $P_{star}$$>$0.5 and $P_{star}$$<$0.5. Sources more likely to be stars concentrate in a well defined arc (the stellar locus) while candidate galaxies occupy a much wider area. The distribution of the $g$-$r$ and $r$-$i$ colours for the spectroscopically confirmed galaxies is roughly consistent with that of the $P_{star}$$<$0.5 subsample.

\begin{figure} 
\begin{center}
\includegraphics[width=8.4cm]{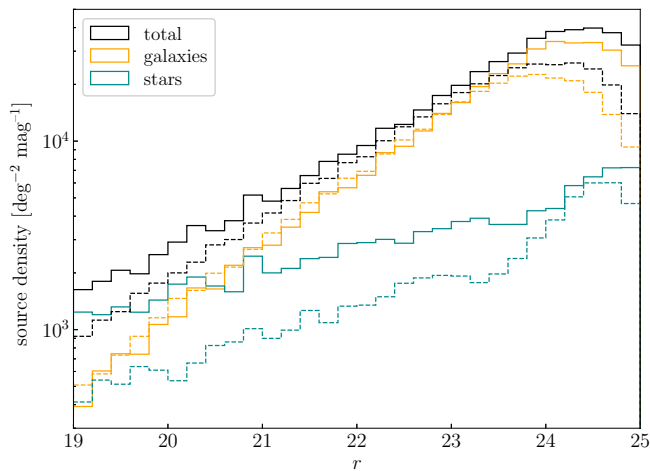}
\end{center}
\caption[]{Density of sources detected per unit area and magnitude as a function of their $r$-band magnitude in J-NEP (solid lines) and miniJPAS (dashed lines). The distributions are shown for all sources (black) as well as separately for galaxies (orange) and stars (cyan) using the probabilities $P_{gal}$ and $P_{star}$ as weights.\label{fig:source-density-mag}}
\end{figure} 

Figure \ref{fig:source-density-mag} compares the density of sources per unit area as a function of their $r$-band magnitude in J-NEP and miniJPAS. The density is computed assuming an effective footprint area (which excludes masked regions) of 0.23 deg$^2$ for J-NEP and 0.895 deg$^2$ for miniJPAS. 
The distribution peaks at $r$$\sim$24.5 in J-NEP compared to $r$$\sim$24.0 in miniJPAS due to deeper $r$-band observations. The density
of sources is higher in J-NEP than in miniJPAS at all magnitudes, but especially at brighter ones.
We weight the individual sources with their $P_{star}$ and $P_{gal}$ = 1 - $P_{star}$ probabilities to produce the distributions for stars and galaxies separately. The density of stars in J-NEP is a factor of $\sim$2 higher relative to miniJPAS due to its lower Galactic latitude. This suffices to explain the excess density of sources in J-NEP relative to miniJPAS. The density of galaxies is consistent between the two surveys for the entire range of magnitudes up to the completeness limit of miniJPAS ($r$$\sim$23.5).
In both surveys there is an upturn in the density of stars at magnitudes fainter than their completeness limit. This is an artefact due to the assignment of $P_{star}$ = 0.5 to sources that are too faint for a morphological classification.

\section{Photometric redshifts}\label{sec:photo-z-main}

\subsection{Computation of photo-$z$\label{sec:photo-z}}

We compute photo-$z$ for all J-NEP sources with magnitude $r$$\le$23 using the same method applied to miniJPAS, which is extensively documented in \citetalias{HC21}. Very briefly, we use a custom version of {\sc LePhare} \citep{Arnouts11} adapted to the properties of the J-PAS data. The galaxy templates are a set of 50 SPS models generated with {\sc CIGALE} \citep{Boquien19}. They were selected to maximise the photo-$z$ performance in a subset of miniJPAS galaxies (see Sect. 4.2 in \citetalias{HC21}). We configure {\sc LePhare} to compute redshift probability distribution functions ($z$PDF) with a resolution of 0.002 in $z$ and a search range 0$<$$z$$<$1.5. A redshift prior derived from galaxy counts in the VIMOS VLT Deep Survey \citep[VVDS;][]{LeFevre05} is applied to the redshift likelihood distributions to obtain the final $z$PDF.

Photo-$z$ are computed for all sources regardless of their morphological classification, but only galaxy templates are used. Therefore, the $zPDF$ obtained by {\sc LePhare} must be considered as redshift probability distribution functions conditional to the source being a galaxy. This allows us to obtain statistics of the galaxy population not biased by the uncertainty in the star/galaxy classification (see Sects. 2.4 and 6 in \citetalias{HC21} for a discussion).

We emphasise that, with the exception of the recalibration strategy, all the other steps in the data processing up to and including the photo-$z$ calculation use the same software packages with the exact same configuration for J-NEP and miniJPAS. Therefore, any statistical differences in the photo-$z$ results must be attributed to one or more of these factors: 
a) differences in the quality/depth of the observations; b) systematics introduced by the recalibration; c) overestimation of the actual photo-$z$ accuracy in miniJPAS due to fine-tuning of the set up in order to maximise the photo-$z$ performance in the spectroscopic subsample of miniJPAS.
In the next sections, we compare the photo-$z$ results for J-NEP and miniJPAS to evaluate the impact of each of these factors. 

\subsection{Distribution of $z_{\rm{phot}}$ and $odds$}

\begin{figure} 
\begin{center}
\includegraphics[width=8.4cm]{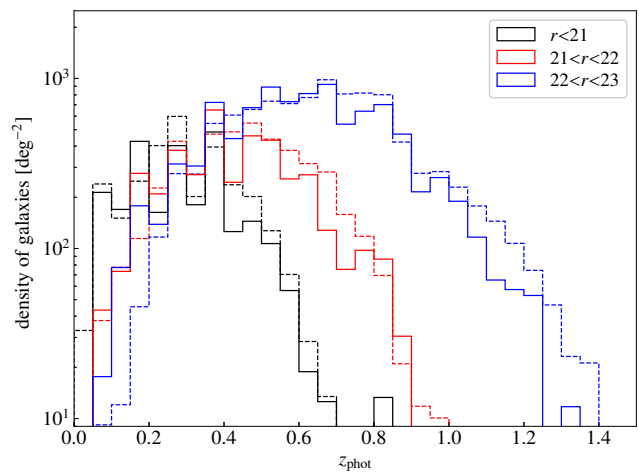}
\end{center}
\caption[]{Density of galaxies per unit area as a function of the most probable photometric redshift ($z_{\rm{phot}}$) for several intervals of $r$-band magnitude in J-NEP (solid lines) and miniJPAS (dashed lines).\label{fig:galaxy-density-zphot}}
\end{figure} 

While the $z$PDF provides the most complete description of our knowledge about the redshift of a source, point estimates can be convenient for some applications. 
The mode of the $zPDF$, $z_{\rm{phot}}$, which represents the redshift with highest probability density, is the most commonly used point estimate.

Another useful scalar is the $odds$, which represents the probability of the actual redshift being within a given interval around the mode. We use an interval of $\pm$3\% in 1+$z$, therefore

\begin{equation}
odds = \int_{z_{\rm{phot}} - d}^{z_{\rm{phot}} + d} P(z) dz, \hspace{1cm} d = 0.03 (1+z_{\rm{phot}})
\end{equation}

Figure \ref{fig:galaxy-density-zphot} compares the distribution of $z_{\rm{phot}}$ for J-NEP and miniJPAS galaxies in three magnitude intervals. As in Fig. \ref{fig:source-density-mag}, all sources are weighted with their corresponding $P_{gal}$ to obtain the redshift distribution for galaxies. 

The distributions of $z_{\rm{phot}}$ for the same magnitude interval are roughly consistent in the two surveys, but there are some interesting small differences. The brightest interval ($r$$<$21) shows high frequency variation with $z_{\rm{phot}}$ in the density of galaxies. For miniJPAS, \citetalias{HC21} showed that the peaks and valleys of the distribution of $z_{\rm{phot}}$ agree with the distribution of spectroscopic redshifts and therefore indicate real over-/under-densities at certain redshifts. In J-NEP, the small survey area implies a higher contrast of the large scale structure (LSS) signature on the redshift distribution. As expected, the over-densities traced by peaks in the distribution of $z_{\rm{phot}}$ occur at different redshifts in J-NEP and miniJPAS, but the overall shape of the distribution is very similar.
The exception is the faintest interval (22$<$$r$$<$23), which shows an excess of sources at $z_{\rm{phot}}$$<$0.2 and a deficit at $z_{\rm{phot}}$$>$1 in J-NEP relative to miniJPAS.
While we cannot rule out cosmic variance affecting the low redshift volume probed by J-NEP due to the small survey area, the deficiency of sources at high redshifts would require a strong under-density within a huge volume of the Universe. A more likely interpretation for both discrepancies is the better photo-$z$ accuracy for 22$<$$r$$<$23 sources in J-NEP relative to miniJPAS due to deeper observations. 
For such faint sources, the low S/N in the photometry (especially in the narrow bands) implies that the redshift prior has a strong effect in the shape of the $zPDF$ and the highest probability density value. 
In particular, for $i$=22 ($r$$\sim$22.5) galaxies the probability density of the prior peaks at $z$$\sim$0.6--0.7 (depending on the spectral type) with a steep decrease at lower $z$ but a flatter one at higher $z$ (see Fig. 7 in \citetalias{HC21}). The very low prior probability discourages $z$$<$0.2 solutions for faint galaxies. 
A higher S/N in J-NEP implies easier detection of spectral features by the template-fitting algorithm, which can help at overcoming the prior in the few faint galaxies that are actually at low $z$.
In contrast, the flatter slope of the high-$z$ tail of the prior is less successful at preventing wrong high-$z$ solutions. In fact, Fig. 14 in \citetalias{HC21} shows that most $z_{\rm{phot}}$$>$1 galaxies in miniJPAS have $z_{\rm{spec}}$$<$1. The higher S/N in J-NEP implies that fewer $z_{\rm{spec}}$$<$1 galaxies are scattered by photometric errors into the $z_{\rm{phot}}$$>$1 region.

\begin{figure} 
\begin{center}
\includegraphics[width=8.4cm]{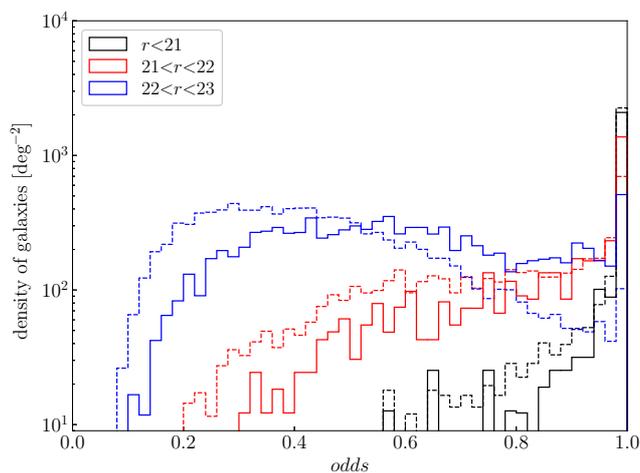}
\end{center}
\caption[]{Density of galaxies per unit area as a function of the $odds$ parameter for several intervals of $r$-band magnitude in J-NEP (solid lines) and miniJPAS (dashed lines).\label{fig:galaxy-density-odds}}
\end{figure}

The higher S/N of the J-NEP photometry relative to miniJPAS at the same magnitude implies higher contrast in the $zPDF$, therefore higher $odds$, for any magnitude interval (Fig. \ref{fig:galaxy-density-odds}). All else being equal, this should translate into a lower outlier rate for J-NEP at all magnitudes. However, the larger systematic uncertainty in J-NEP colours implies that, in sources bright enough for systematic errors to dominate over photometric errors, the outlier rate in J-NEP can be higher than in miniJPAS (see Sect. \ref{sec:validation}).

\section{Accuracy of photo-$z$ in J-NEP}\label{sec:accuracy}

\subsection{Validation with the spectroscopic sample}\label{sec:validation}

\begin{figure} 
\begin{center}
\includegraphics[width=8.7cm]{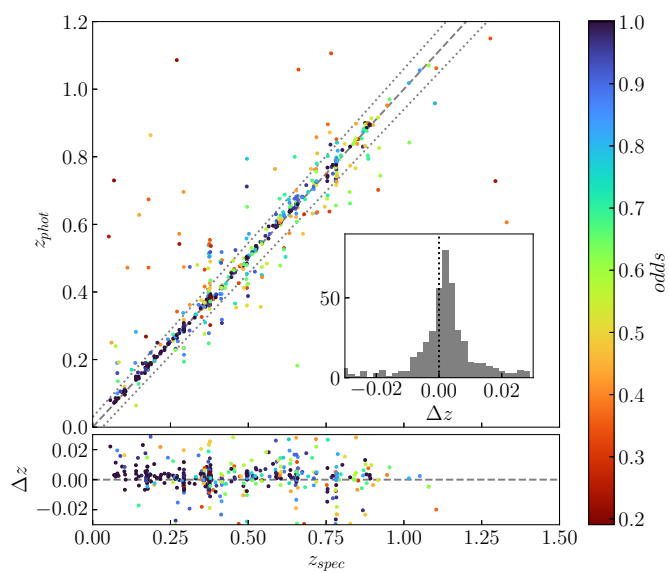}
\end{center}
\caption[]{Photometric redshift (top panel) and photo-$z$ error (bottom panel) versus spectroscopic redshift for individual sources in J-NEP. Only $r$$<$23 sources with no photometry flags (FLAGS = 0 and MASK\_FLAGS = 0) and reliable $z_{\rm{spec}}$ are shown. The colour coding indicates the $odds$ parameter that represents the confidence in the $z_{\rm{phot}}$ solution. The dashed line indicates the $z_{\rm{phot}}$=$z_{\rm{spec}}$ relation. Dotted lines in the top panel enclose the region corresponding to $\vert\Delta z\vert$$<$0.03, which is enlarged in the bottom panel. The inset plot shows the distribution of $\Delta z$.\label{fig:zphot-zspec}}
\end{figure}

\begin{figure} 
\begin{center}
\includegraphics[width=8.4cm]{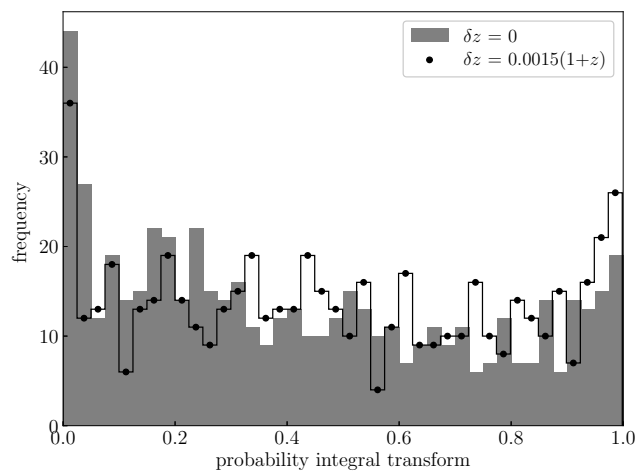}
\end{center}
\caption[]{Distribution of PIT values for the $z$PDFs of J-NEP galaxies with spectroscopic redshift. The solid and open histograms represent the distributions before and after correcting for a systematic offset $\delta z$ = 0.0015(1+$z$) in the $z$PDFs.\label{fig:PIT-diagram}}
\end{figure}

\begin{figure} 
\begin{center}
\includegraphics[width=8.4cm]{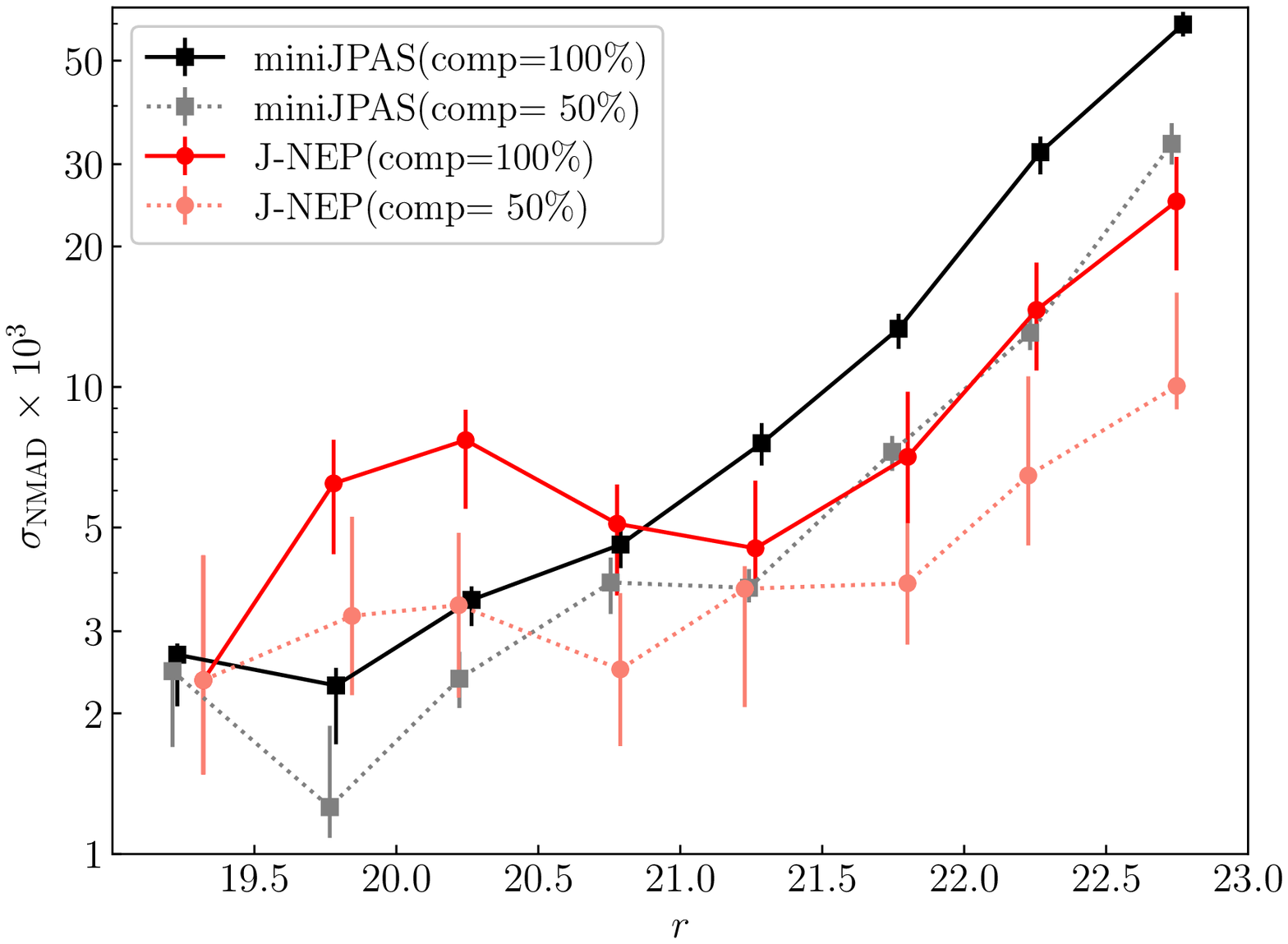}
\includegraphics[width=8.4cm]{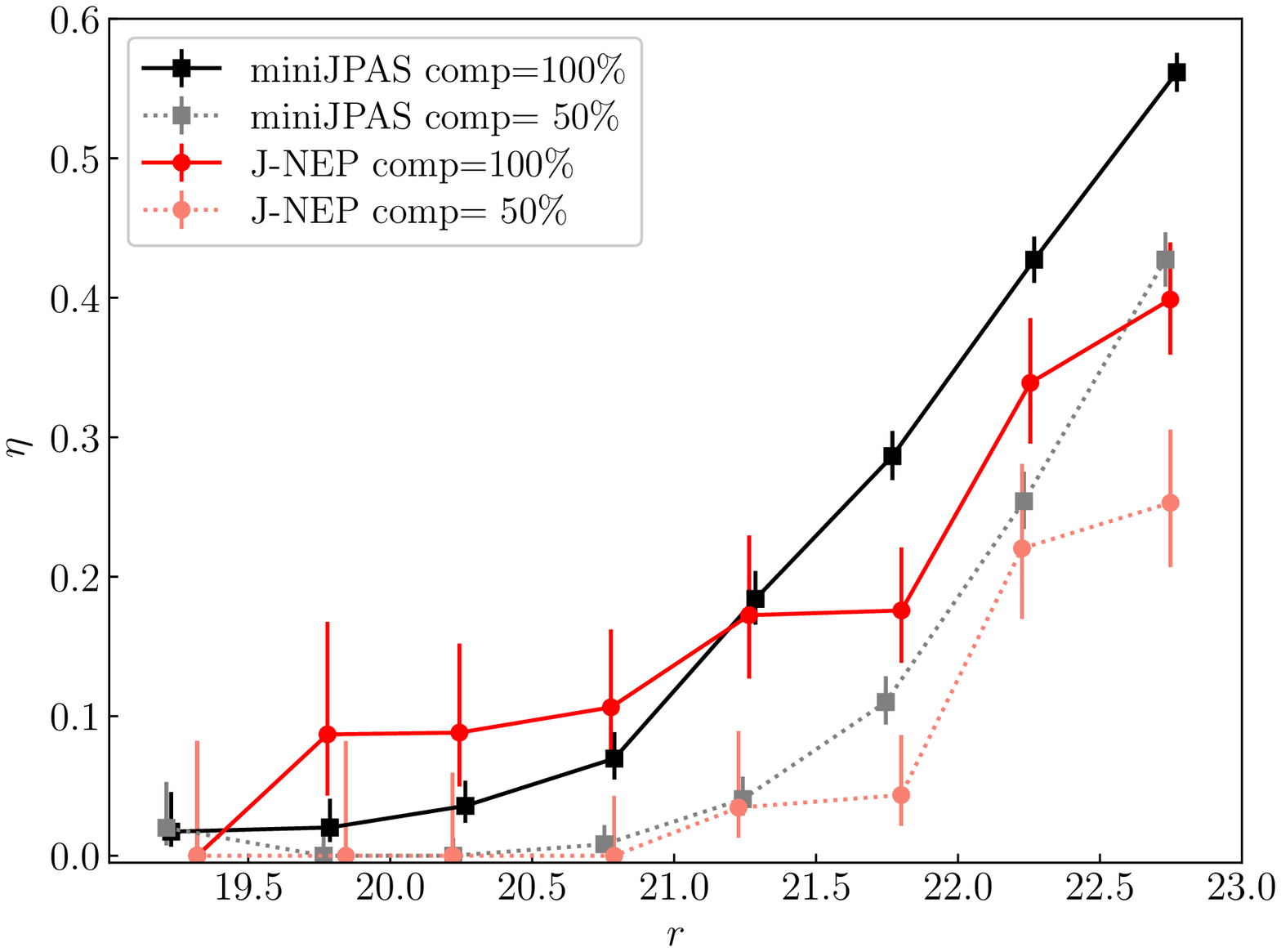}
\end{center}
\caption[]{Normalised median absolute deviation of the photo-$z$ error ($\sigma_{\rm{NMAD}}$; top panel) and outlier rate ($\eta$; bottom panel) in bins of $r$-band magnitude for the spectroscopic subsamples of miniJPAS (black squares) and J-NEP (red circles). Symbols linked by solid lines represent measurements including all sources within a 0.5 mag interval in $r$-band magnitude, while the ones linked by dotted lines indicate the values obtained considering only the 50\% of sources with higher $odds$ in each magnitude bin. Error bars indicating the 16--84$^{th}$ percentile confidence intervals were obtained with bootstrap resampling in the case of $\sigma_{\rm{NMAD}}$ and the Wilson formula for binomial distributions in the case of $\eta$.\label{fig:snmad-by-mag}}
\end{figure}

In this section, we use the subsample of J-NEP with
spectroscopic redshifts to evaluate the accuracy of photo-$z$ estimates.

Figure \ref{fig:zphot-zspec} compares $z_{\rm{phot}}$ and $z_{\rm{spec}}$ values for the 554 J-NEP galaxies with reliable spectroscopic redshifts in the catalog from Willmer et al. (in preparation) and no flags in their photometry.
The inset plot shows the distribution of the relative error in $z_{\rm{phot}}$, defined as $\Delta z$ = ($z_{\rm{phot}}$ - $z_{\rm{spec}}$)/(1+$z_{\rm{spec}}$).

The distribution of $\Delta z$ shows that $z_{\rm{phot}}$ systematically overestimates the actual redshift of the galaxies by a small amount (median($\Delta z$) $\sim$ 0.1--0.2\%, depending on the magnitude and $odds$ cut applied and the algorithm used to reject outliers). This is consistent with the median($\Delta z$) $\sim$ 0.10--0.14\% found in miniJPAS \citepalias{HC21}.
The probability integral transform (PIT) diagram (Fig. \ref{fig:PIT-diagram}) shows that, excluding PIT values close to 0 and 1 \citep[which are dominated by catastrophic redshift errors, see e.g.][]{Polsterer16}, the distribution is slightly tilted, suggesting that this bias affects the entire $z$PDF, not just its mode (this was also the case in miniJPAS).
In the following analysis we correct all $z_{phot}$ measurements in J-NEP and miniJPAS by subtracting $\delta z$ = 0.0015(1+$z$).

The distribution of $\Delta z$ is far from Gaussian. \citetalias{HC21} showed that its shape is well reproduced by the combination of two Lorentzian profiles (or a Gaussian plus a Lorentzian) of very different widths. The narrow component corresponds to sources affected by small inaccuracies in the determination of the peak of the $zPDF$, while the broad component is dominated by ``catastrophic errors'' due to a flattened or multi-modal $zPDF$.

The typical error in $z_{\rm{phot}}$ (excluding catastrophic errors) is often represented by the $\sigma_{\rm{NMAD}}$ statistic, defined as

\begin{equation}
\sigma_{\rm{NMAD}} = 1.48 \times \textrm{median}\big{\vert}\Delta z_i - \textrm{median}(\Delta z_i)\big{\vert}
\end{equation}

Another frequently used statistic is the outlier rate

\begin{equation}
\eta = \frac{N(\vert \Delta z \vert > X)}{N_{tot}}
\end{equation}

\noindent where $N_{tot}$ is the number of sources in the sample and $N(\vert \Delta z \vert > X)$ is the number of them having relative errors in $z_{\rm{phot}}$ larger than a threshold $X$. If we choose $X$ = 0.03, then the expectation value of $\eta$ for the sample is related to the average $odds$ by the equation

\begin{equation}
E{[}\eta{]} = 1 - \langle odds \rangle
\end{equation}
While the majority of the sources in Fig. \ref{fig:zphot-zspec} are tightly packed near the 1:1 relation, 132 sources (24\% of the sample) are outside the region enclosed by the dotted lines that represent an error $\vert\Delta z\vert$ = 0.03. This 24\% of outliers is consistent with the $\sim$23\% predicted from the mean value of their $odds$ parameter, $\langle$$odds$$\rangle$ = 0.771. 

\begin{figure} 
\begin{center}
\includegraphics[width=8.4cm]{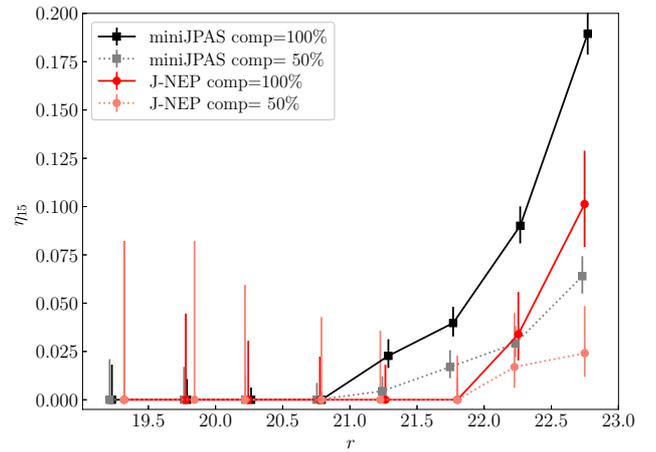}
\end{center}
\caption[]{Frequency of extreme outliers ($\vert\Delta z\vert$ $>$ 0.15) in bins of $r$-band magnitude for the spectroscopic subsamples of miniJPAS and J-NEP. Symbols as in Fig. \ref{fig:snmad-by-mag}.\label{fig:outrate-by-mag-eta15}}
\end{figure}

Figure \ref{fig:snmad-by-mag} compares $\sigma_{\rm{NMAD}}$ and $\eta$ in bins of magnitude for J-NEP and miniJPAS. Values are shown at 100\% completeness (no cut in $odds$ applied) and 50\% completeness (only 50\% of sources with higher $odds$ in each magnitude bin are used to compute $\sigma_{\rm{NMAD}}$ and $\eta$). We note that $\eta$ values for miniJPAS and J-NEP are high in part because of our restrictive criterion for outliers. Using the criterion $\vert \Delta z \vert$ $>$ 0.15 that is typical for broadband surveys results in much lower outlier rates (Fig. \ref{fig:outrate-by-mag-eta15}). In particular, none of the J-NEP galaxies with $r$$<$22 and no flags has $\vert \Delta z \vert$ $>$ 0.15.

The error bars for $\sigma_{\rm{NMAD}}$ and $\eta$ in J-NEP are large due to the relatively small spectroscopic sample. In spite of this, we find some interesting trends.
For sources fainter than $r$=21.5 both $\sigma_{\rm{NMAD}}$ and $\eta$ are $\sim$30--50\% lower in J-NEP compared to miniJPAS at the same magnitude and completeness. This is probably a consequence of the deeper images of J-NEP and their narrower PSF, which allows for higher S/N in the photometry, in particular in the red bands (see Fig. \ref{fig:FWHM-depth}).

At magnitudes brighter than $r$=21 the values of $\sigma_{\rm{NMAD}}$ and $\eta$ are higher in J-NEP compared to miniJPAS. This suggests that for bright sources the photo-$z$ accuracy is less sensitive to the S/N of the photometry as other factors become dominant.
In particular, larger systematic errors in the recalibration with the galaxy locus relative to ISMFR implies that the total flux errors (systematic plus photometric) can be higher for bright J-NEP sources compared to miniJPAS sources of the same magnitude. 
However, systematics in the recalibration cannot explain a higher $\sigma_{\rm{NMAD}}$ at $r$$\sim$20 than at $r$$\sim$21 for J-NEP. This puzzling result might be just a consequence of small number statistics ($\sigma_{\rm{NMAD}}$ values for the intervals between $r$=19.5 and $r$=21.5 agree within their uncertainties). We find no clear difference in the properties of $r$$\sim$20 and $r$$\sim$21 galaxies that could explain this trend and propose further investigation when a larger spectroscopic sample becomes available.
Irrespective of what is the actual photo-$z$ accuracy at the bright end, the steep increase in the density of sources at fainter magnitudes implies that J-NEP has better overall accuracy than miniJPAS.

\begin{figure} 
\begin{center}
\includegraphics[width=8.4cm]{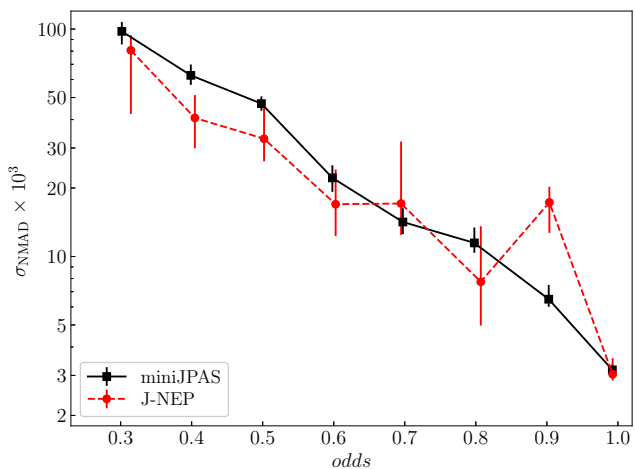}
\end{center}
\caption[]{$\sigma_{\rm{NMAD}}$ computed in bins of $odds$ of width 0.1 for the spectroscopic sources in miniJPAS (squares) and J-NEP (circles). Error bars represent 1-$\sigma$ confidence intervals obtained with bootstrap resampling.\label{fig:snmad-by-odds}}
\end{figure}

\citetalias{HC21} showed that the $odds$ parameter is a reliable predictor of the probability of a redshift outlier and its correlation with $\vert\Delta z\vert$ is stronger compared to other properties such as the $r$-band magnitude. They also showed that the $\sigma_{\rm{NMAD}}$ within a small interval of $odds$ does not have any clear residual dependence with the $r$-band magnitude, $z_{\rm{spec}}$, or the spectral type.
In Fig. \ref{fig:snmad-by-odds}, we compare the dependence of $\sigma_{\rm{NMAD}}$ with $odds$ in the spectroscopic samples of miniJPAS and J-NEP. Values of $\sigma_{\rm{NMAD}}$ at the same $odds$ agree within their 1-$\sigma$ uncertainties for all intervals except those centred at $odds$=0.4 and $odds$=0.9 (the excess $\sigma_{\rm{NMAD}}$ at $odds$$\sim$0.9 in J-NEP also deserves further investigation), despite the differences in depth and recalibration methods. This suggests that the relation between $odds$ and $\sigma_{\rm{NMAD}}$ is largely insensitive to a change of $\sim$0.5 mag in the depth of the observations and is not strongly affected by the systematic uncertainties resulting from the different recalibration strategies in J-NEP and miniJPAS.

Because of the inevitable small differences in the PSF FWHM and/or depth of the coadded J-PAS images, the S/N of the sources at a given magnitude (and therefore also their photo-$z$ accuracy) will vary between pointings. However, since these changes in the quality of the photometry are automatically accounted for by the $odds$ parameter, we expect the $\sigma_{\rm{NMAD}}$ and $\eta$ at constant $odds$ to remain homogeneous throughout the J-PAS survey.

\subsection{Extrapolation to the entire J-NEP}

\begin{figure} 
\begin{center}
\includegraphics[width=8.4cm]{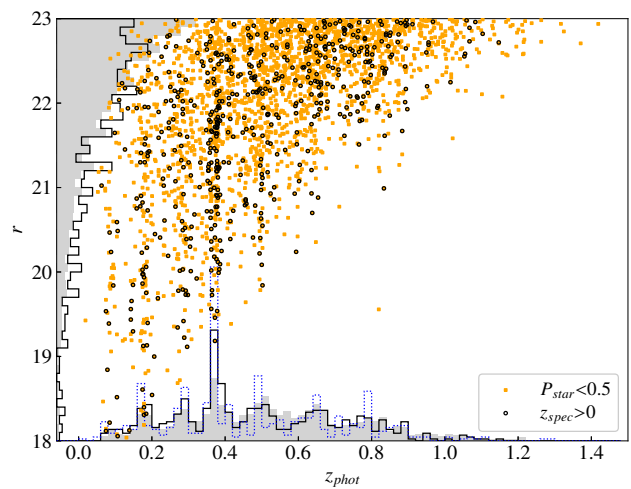}
\end{center}
\caption[]{Magnitude in the $r$ band and the highest probability photometric redshift for individual galaxies ($P_{star}$$<$0.5) in J-NEP. Black circles indicate sources with spectroscopic redshift. The solid grey and step black line histograms represent the $r$-band magnitude and $z_{\rm{phot}}$ distributions for all galaxies and galaxies with spectroscopic redshift, respectively. The blue dotted histogram represents the distribution of spectroscopic redshifts in the spectroscopic subsample.\label{fig:rmag-photoz}}
\end{figure}

The subsamples with and without spectroscopy in miniJPAS have very similar distributions for the $r$-band magnitude, broad-band colours, $z_{\rm{phot}}$ and $odds$. This allowed \citetalias{HC21} to conclude that the photo-$z$ performance in the two subsamples must also be very similar. 
In the case of J-NEP, we have already shown in Fig. \ref{fig:gri-star-galaxy} that the distribution of the $g$-$r$ and $r$-$i$ colours in the spectroscopic sample is roughly consistent with that of the entire sample of $P_{star}$$<$0.5 sources. However, both the $r$-band magnitude and the photo-$z$ show highly significant differences in their distributions (see Fig. \ref{fig:rmag-photoz}). In particular, only 24\% of galaxies in the spectroscopic sample have $r$$>$22.5 
compared to 35\% in the full sample. On the other hand, bright galaxies ($r$$<$21) represent 25.5\% of the spectroscopic sample but just 17\% of the full sample. This bias towards brighter galaxies is also reflected in the distribution of $z_{\rm{phot}}$: 57\% of the spectroscopic sample has $z_{\rm{phot}}$$<$0.5 compared to just 47\% in the full sample.

Since the accuracy of photo-$z$ estimates is sensitive to both the magnitude and redshift of the sources (which determine the S/N of the photometry and the spectral features covered by the observed wavelength range, respectively), we can expect that the $\sigma_{\rm{NMAD}}$ and $\eta$ measured in the spectroscopic subsample of J-NEP underestimate their values in the full sample.

To predict the distribution of $\vert\Delta z\vert$ in the full miniJPAS sample, \citetalias{HC21} compensated for the small statistical differences between the samples with and without $z_{\rm{spec}}$ in miniJPAS by modelling the latter using a random subset of the former with the same distribution of $odds$. Since no significant dependence of $\sigma_{\rm{NMAD}}$ or $\eta$ with other parameters at constant $odds$ was found, it was reasonable to assume that two samples with the same $odds$ distribution also share a similar distribution of $\vert\Delta z\vert$.
While the spectroscopic subsample of J-NEP is too small to check for variation in the photo-$z$ accuracy at constant $odds$, the experience with miniJPAS suggests that such dependence should be small. Therefore, we can predict the distribution of $\vert\Delta z\vert$ (and from it $\sigma_{\rm{NMAD}}$ and $\eta$) by substituting each of the sources in the full sample with one from the spectroscopic subsample that has similar $odds$\footnote{In practice, we define a small tolerance, $\Delta$$odds$$=$0.005, and pick at random one of the sources within $odds$$\pm$$\Delta$$odds$.}.
We repeat this source association by $odds$ multiple times. The randomness of the process ensures that in each realisation a different subset of the spectroscopic sample is selected and therefore the resulting distribution of $\Delta z$ changes. This provides an indication of the statistical uncertainty in the predicted distribution of $\Delta z$ for the full sample.

\begin{figure} 
\begin{center}
\includegraphics[width=8.4cm]{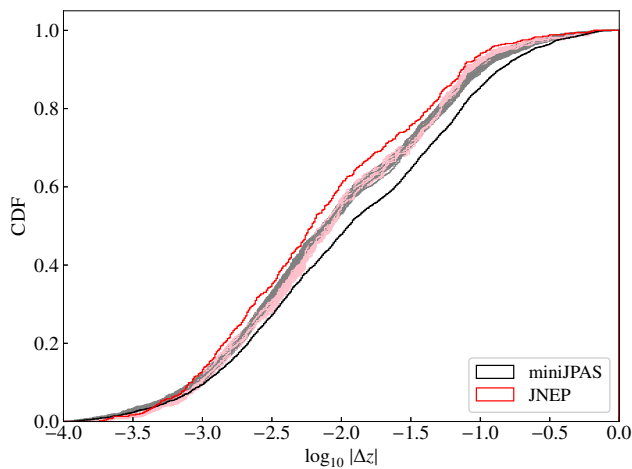}
\end{center}
\caption[]{Cumulative distribution function of photo-$z$ errors in the spectroscopic samples of J-NEP (red) and miniJPAS (black), and predictions for all J-NEP galaxies based on the $odds$-matching with the spectroscopic samples of J-NEP (pink) and miniJPAS (grey).\label{fig:dz-distrib-predict}}
\end{figure}

The results are shown in Fig. \ref{fig:dz-distrib-predict}. The cumulative distribution functions (CDFs) of $\vert\Delta z\vert$ for each of the realisations (pink histograms) are similar in shape, but systematically below the CDF for the spectroscopic sample (red histogram). The predicted $\sigma_{\rm{NMAD}}$ for the full sample is 1.16$\pm$0.07 $\times$ 10$^{-2}$ compared to 0.88 $\times$ 10$^{-2}$ in the spectroscopic sample. For the outlier rate, the prediction is $\eta$=30.0$\pm$0.7\% against $\eta$=25.0\% in the spectroscopic sample.

We can also predict the distribution of $\vert\Delta z\vert$ in J-NEP with a match by $odds$ to the spectroscopic sample in miniJPAS instead of that in J-NEP. The resulting CDFs are shown as grey histograms in Fig. \ref{fig:dz-distrib-predict}. They overlap with the spectroscopic sample of J-NEP in most of the range, but there are small discrepancies at the extremes ($\log_{
\rm{10}}$ $\vert\Delta z\vert$ $\lesssim$-3 and $\log_{
\rm{10}}$ $\vert\Delta z\vert$ $\gtrsim$-1.2) and also around $\log_{\rm{10}}$ $\vert\Delta z\vert$ $\sim$ -2.5. However, the predictions for $\sigma_{\rm{NMAD}}$ (1.13$\pm$0.05 $\times$ 10$^{-2}$) and $\eta$ (29.0$\pm$0.7\%) are both consistent within the uncertainties with those from the spectroscopic sample of J-NEP.

Obtaining compatible predictions from the spectroscopic samples in J-NEP and miniJPAS despite the differences in the depth of the images implies that we will be able to evaluate the photo-$z$ performance in individual J-PAS pointings from their $odds$ distribution alone, regardless of the availability of spectroscopy. 
 
\subsection{Photo-$z$ accuracy in magnitude- and $odds$-limited samples}

\begin{figure} 
\begin{center}
\includegraphics[width=8.4cm]{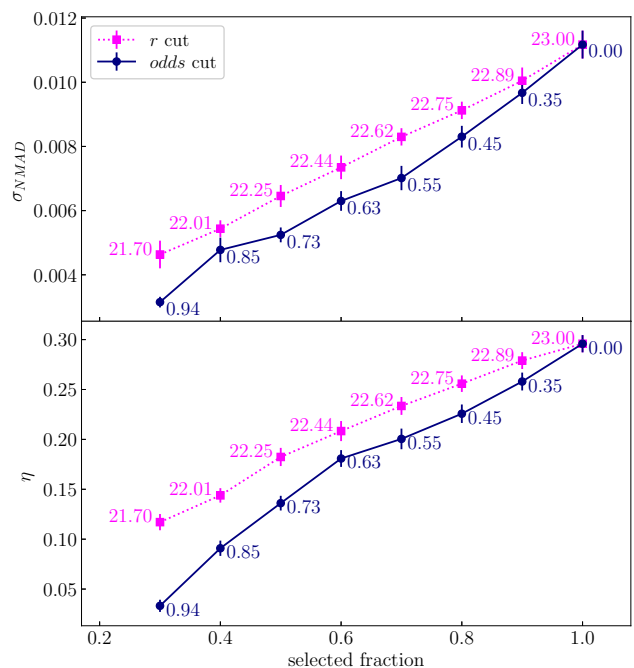}
\end{center}
\caption[]{Dependence of $\sigma_{\rm{NMAD}}$ (top panel) and $\eta$ (bottom panel) on the fraction of the full J-NEP sample selected when a threshold value in the $r$-band magnitude (squares) or the $odds$ parameter (circles) is applied. Each symbol is annotated with the numerical value of the threshold.\label{fig:snmad-outrate-completeness}}
\end{figure}

In the previous section, we estimated that the photo-$z$ of J-NEP sources have typical errors of $\sim$1.1\%, with a $\sim$30\% probability for the error being larger than 3\%. While this may suffice for some applications, others require higher accuracy (lower $\sigma_{\rm{NMAD}}$), higher reliability (lower $\eta$), or both. 
This can be achieved, at the cost of sample size, by limiting the selection in magnitude or in $odds$. While a cut in magnitude seems more natural and easier to work with, \citetalias{HC21} showed that cutting in $odds$ is significantly more efficient.
In Fig. \ref{fig:snmad-outrate-completeness}, we show $\sigma_{\rm{NMAD}}$ and $\eta$ predicted for a selection of the full J-NEP sample in which a threshold value in $odds$ or the $r$-band magnitude is applied. At any given selection fraction, both $\sigma_{\rm{NMAD}}$ and $\eta$ are consistently lower for an $odds$-based selection. Also, if the goal is to reach some specific value of  $\sigma_{\rm{NMAD}}$ or $\eta$, the fraction of sources selected is consistently larger with an $odds$-based selection. 

\section{On the nature of photo-$z$ outliers}\label{sec:outliers}

\citetalias{HC21} proposed three possible reasons for photo-$z$ outliers: degeneracy in the colour space, artefacts in the photometry, and exotic spectra not represented in the template set. 
In this section we explore in more detail the factors contributing to photo-$z$ outliers using the spectroscopic samples of miniJPAS and J-NEP.

\begin{figure} 
\begin{center}
\includegraphics[width=8.4cm]{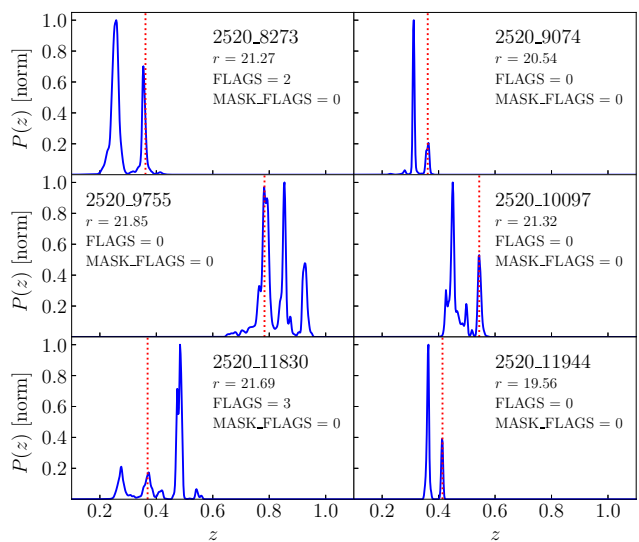}
\end{center}
\caption[]{Examples of $z$PDFs for J-NEP sources in which the spectroscopic redshift (dotted line) matches a secondary peak of the $z$PDF.\label{fig:pdf-double-peak}}
\end{figure}

Colour-space degeneracy occurs when the observed photometry of a source can be reproduced with comparably good accuracy (i.e., similar $\chi^2$) by two or more spectral templates at significantly different redshifts. This translates into a likelihood distribution $\mathcal{L}$($z$) $\propto$ $e^{-\chi^2(z)/2}$ with two or more peaks, only one of them matching the actual redshift of the source. Photometric errors and artefacts in the photometry can in some cases induce a better fit (lower $\chi^2$) at the  wrong redshift, resulting in a photo-$z$ outlier (see Figure \ref{fig:pdf-double-peak}).

The redshift prior modulates $\mathcal{L}$($z$) by penalising values of $z$ that imply a very high or very low luminosity at a given magnitude. For bright sources, the resulting $P$($z$) sees a decrease in the strength of peaks at high $z$ compared to $\mathcal{L}$($z$). For fainter ones, peaks at low $z$ are suppressed. 
While this decreases the frequency of spurious solutions in normal galaxies, it also makes harder to select the correct peak for low $z$ sources with unusually low luminosity, like dwarf galaxies. 

\begin{figure} 
\begin{center}
\includegraphics[width=8.4cm]{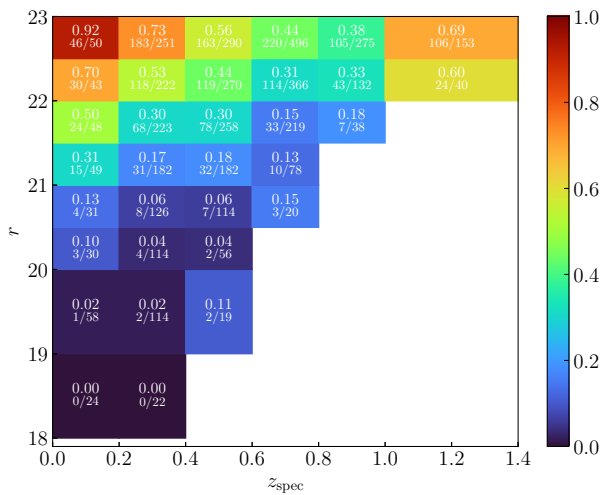}
\end{center}
\caption[]{Colour-coded map of the outlier rate as a function of $r$-band magnitude and redshift for the combined spectroscopic samples of miniJPAS and J-NEP. The colour of each rectangular region represents the frequency of outliers ($\vert\Delta z\vert$$>$0.03) among sources inside that region.\label{fig:mag-z-outliers}}
\end{figure}

\begin{figure} 
\begin{center}
\includegraphics[width=8.4cm]{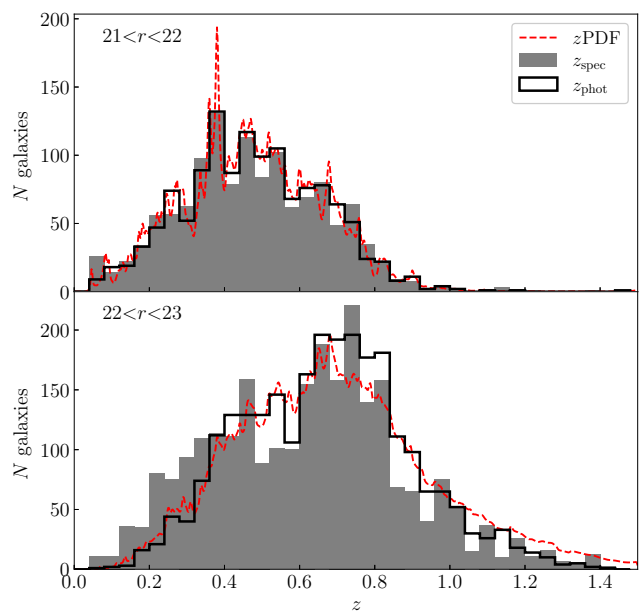}
\end{center}
\caption[]{Redshift distribution for galaxies in the combined spectroscopic samples of miniJPAS and J-NEP for the magnitude bins 21$<$$r$$<$22 (top) and 22$<$$r$$<$23 (bottom). The grey bars and solid black lines indicate the distribution of $z_{\rm{spec}}$ and $z_{\rm{phot}}$, respectively. The dashed red line represent the combined $z$PDFs for the galaxies in the magnitude bin.\label{fig:pdf-redshift}}
\end{figure}

The impact of the prior in the outlier rate is evident in Figure \ref{fig:mag-z-outliers}. 
For each magnitude bin, the minimum $\eta$ is obtained at a redshift close to the peak probability of the prior for that magnitude \citepalias[see Fig. 7 in][]{HC21}. The effect is most dramatic in the faintest magnitude bin ($r$=22.5--23), where the outlier rate increases from $\eta$=0.38 at $z$=0.8--1 to $\eta$=0.92 at $z$=0--0.2. 
The dependence of $\eta$ with $z$ at constant magnitude is an expected consequence of having a redshift prior and does not imply that the prior or the $z$PDFs are biased. The top panel in Figure \ref{fig:pdf-redshift} shows that for the 21$<$$r$$<$22 range, the distribution of $z_{\rm{phot}}$ agrees with that for $z_{\rm{spec}}$ as well as with the combined $z$PDF of all the galaxies in that magnitude range (in other words, taking the mode of the $z$PDF does not bias the results against low redshift galaxies). However, in the magnitude range 22$<$$r$$<$23 both the combined $z$PDF and the distribution of $z_{\rm{phot}}$ predict far fewer galaxies at $z$$<$0.36 than observed. This suggests the prior might underestimate their number density, albeit a selection effect in the spectroscopic sample might also explain the discrepancy (H$\alpha$ exits the spectroscopic range at $z$$\sim$0.38, making more difficult to obtain a high confidence $z_{\rm{spec}}$ at faint magnitudes).

Figure \ref{fig:mag-z-outliers} also shows a sudden increase in $\eta$ at $z$$>$1 compared to the $z$=0.8--1 interval. The redshift prior is unlikely to be responsible for this since it decreases smoothly towards higher $z$. A more likely interpretation is the shift outside of the J-PAS wavelength range of another useful spectral feature, the 4000 \AA{} break, at $z$$\sim$1.2. While the [O {\sc ii}] 3727 \AA{} line allows for reliable $z_{\rm{spec}}$ up to $z$$\sim$1.4, in most cases it is not strong enough for detection in the reddest narrow band filters of J-PAS.

\begin{figure} 
\begin{center}
\includegraphics[width=8.4cm]{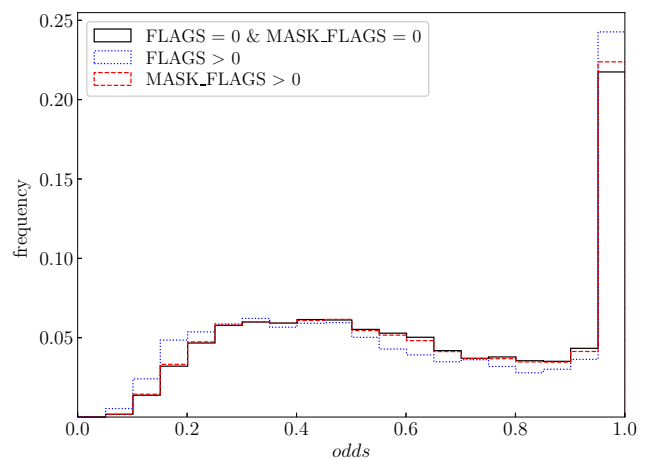}
\end{center}
\caption[]{Distribution of $odds$ for galaxies in the combined sample of miniJPAS and J-NEP sources with $r$$<$23, separating flagged and non-flagged sources using the FLAGS and MASK\_FLAGS parameters (see text for details). Each source is weighted with its $P_{gal}$.\label{fig:odds-distrib-flags}}
\end{figure}

The second proposed cause for outliers, artefacts in the photometry, includes any circumstances that introduce errors in the observed photometry not accounted for by the nominal photometric uncertainty. These may include: systematics in the photometric calibration, inaccurate background subtraction, contamination from nearby sources,  saturation, residuals from cosmic rays, defects in the CCD, etc. 

Some of these issues are identified by the data reduction pipeline or {\sc SExtractor} and stored in the MASK\_FLAGS or FLAGS parameters, respectively \citep[see][for details]{Bonoli21}.
MASK\_FLAGS = 1 indicates the source is close to the borders of the image, while sources  affected by saturated stars have MASK\_FLAGS = 2. The most relevant {\sc SExtractor} flags are FLAGS = 1 (photometry likely biased by neighbouring sources or by more than 10\% of bad pixels in any aperture), FLAGS = 2 (the object has been deblended), and FLAGS = 16 (at least one photometric aperture is incomplete or corrupted). The flags are additive. Sources with no detected issues have MASK\_FLAGS = 0 and FLAGS = 0.

The impact of flagged issues in the photo-$z$ is small for most of the sources. Figure \ref{fig:odds-distrib-flags} shows that the distribution of the $odds$ parameter for sources with MASK\_FLAGS $>$ 0 is consistent with that for sources with no flags. Sources with FLAGS $>$ 0, on the other hand, have a higher probability of obtaining either very low ($<$0.25) or very high $odds$ ($>$0.95). 

\begin{figure} 
\begin{center}
\includegraphics[width=8.4cm]{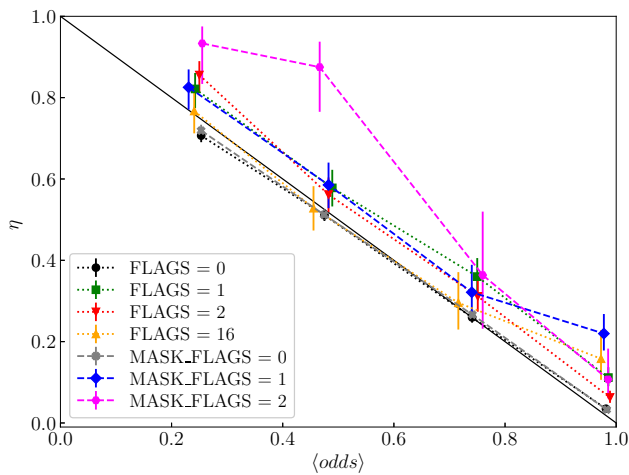}
\end{center}
\caption[]{Relation between the outlier rate and the average $odds$ for sources grouped by their flags values. Symbols indicate the fraction of outliers found in the combined spectroscopic sample of miniJPAS and J-NEP for sources in the $odds$ intervals 0--0.35, 0.35--0.6, 0.6--0.9, and 0.9--1. Error bars represent 68\% confidence intervals computed with the Wilson formula for binomial distributions. The solid line represents the relation $\eta$ = 1 - $\langle odds \rangle$.\label{fig:eta-odds-flags}}
\end{figure}

\begin{figure*} 
\begin{center}
\begin{tikzpicture}
\draw (0, 0) node[inner sep=0] {\includegraphics[height=8.2cm]{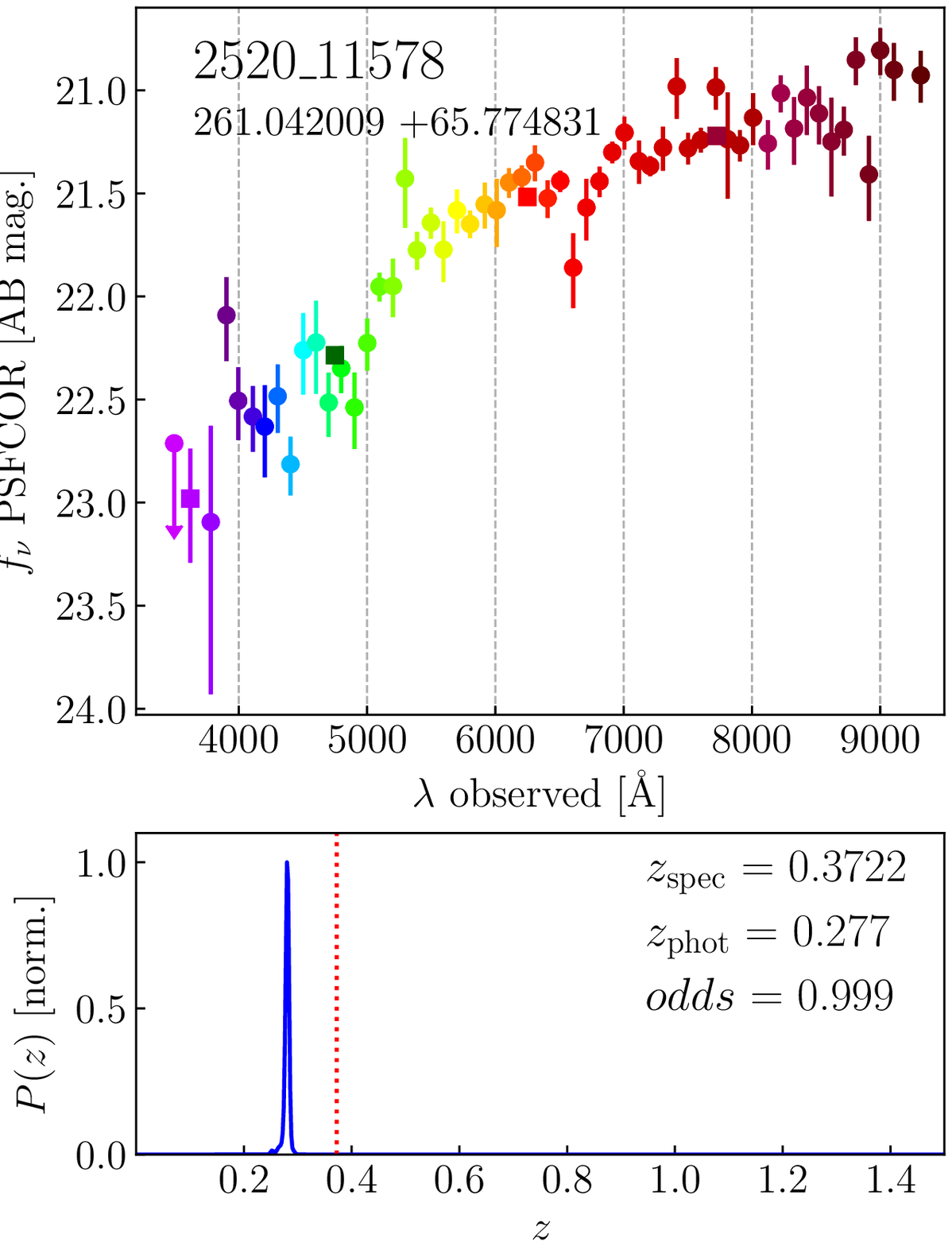}};
\draw (2.25,0.3) node[inner sep=0]{\includegraphics[height=1.7cm]{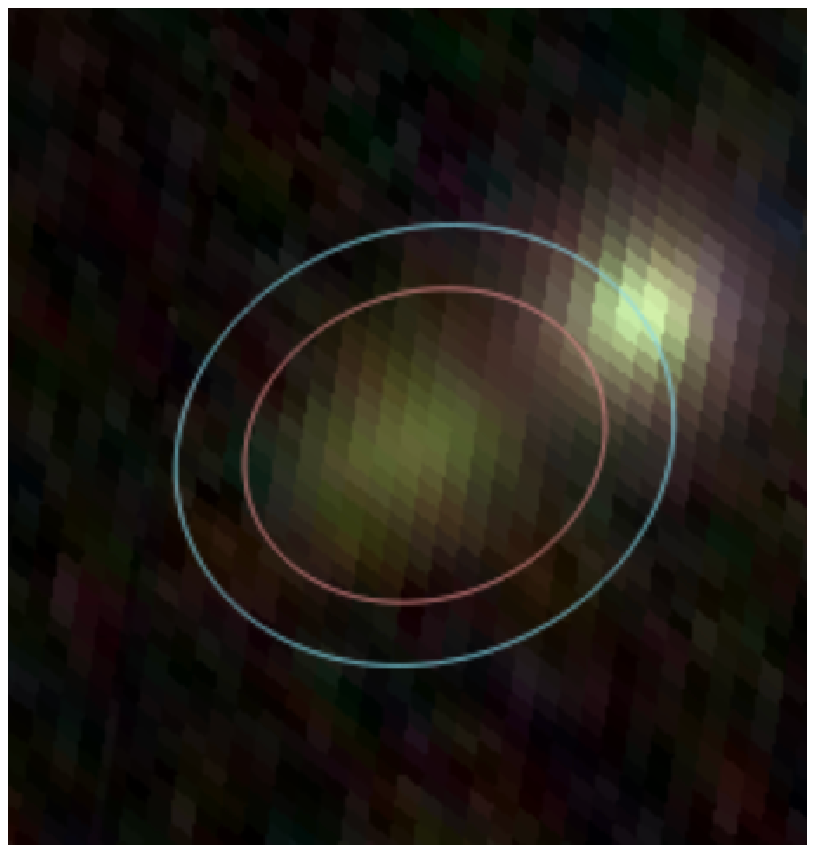}};
\end{tikzpicture}
\begin{tikzpicture}
\draw (0, 0) node[inner sep=0] {\includegraphics[height=8.2cm]{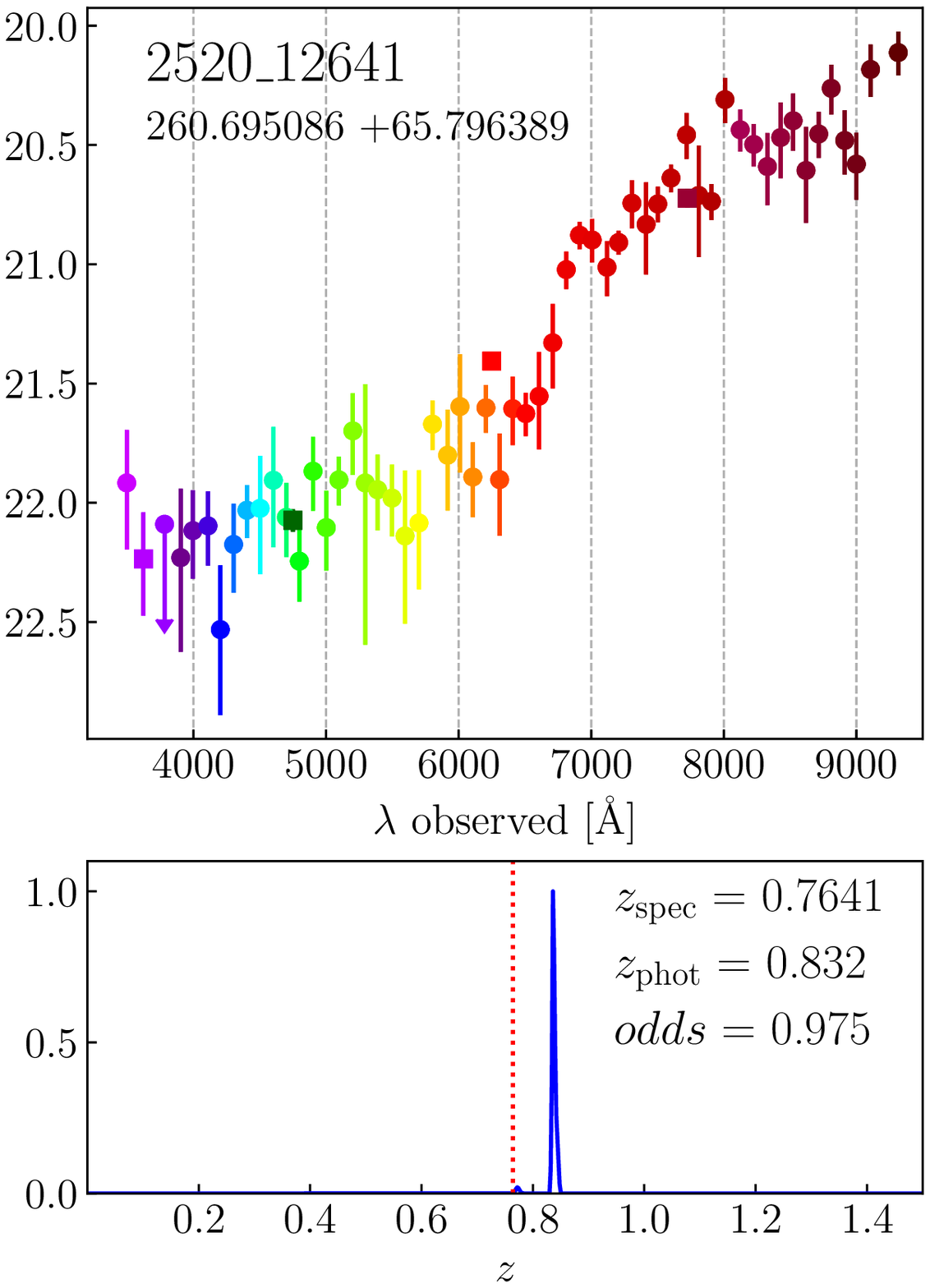}};
\draw (2.05,0.3) node[inner sep=0]{\includegraphics[height=1.7cm]{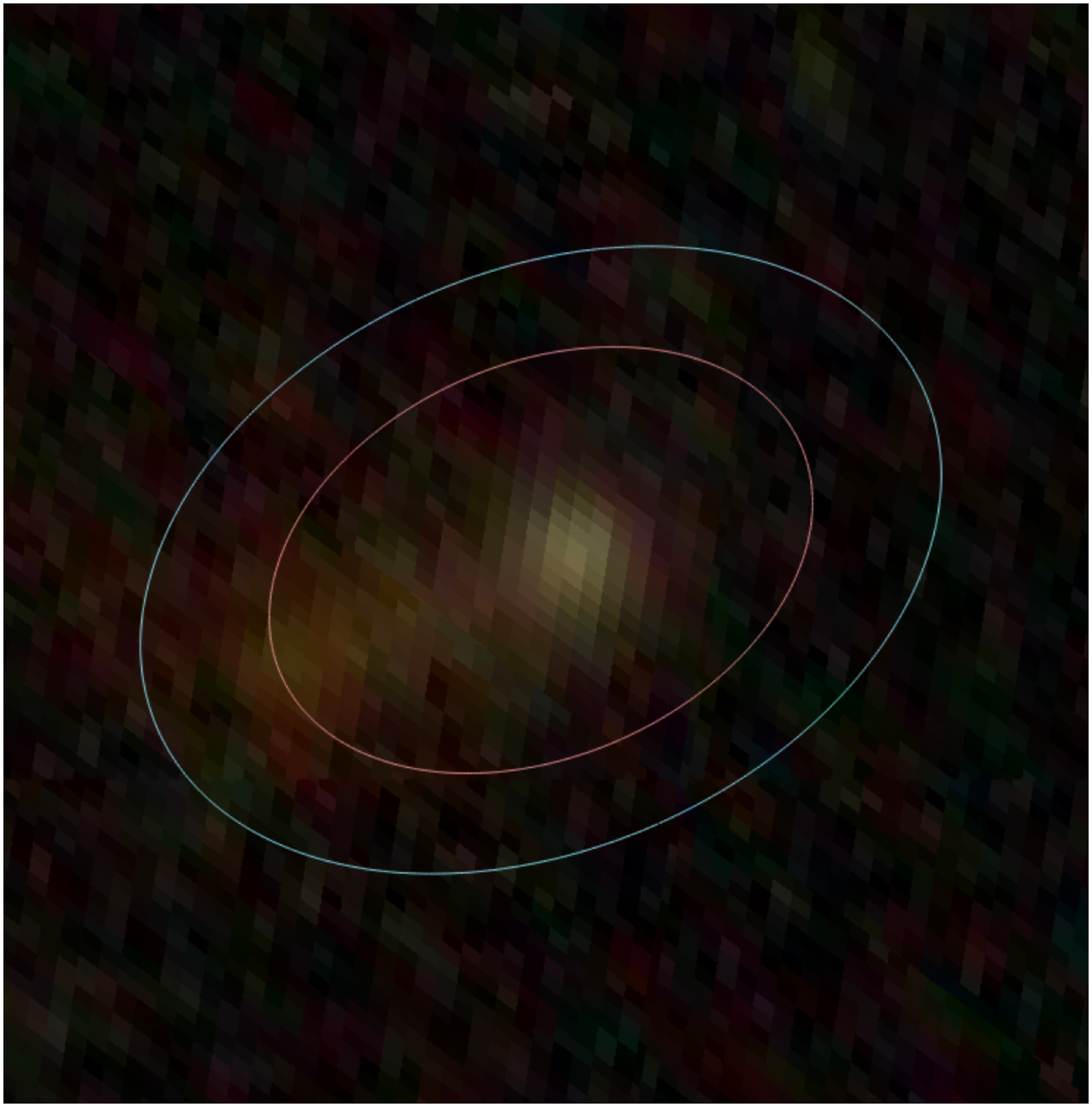}};
\end{tikzpicture}
\begin{tikzpicture}
\draw (0, 0) node[inner sep=0] {\includegraphics[height=8.2cm]{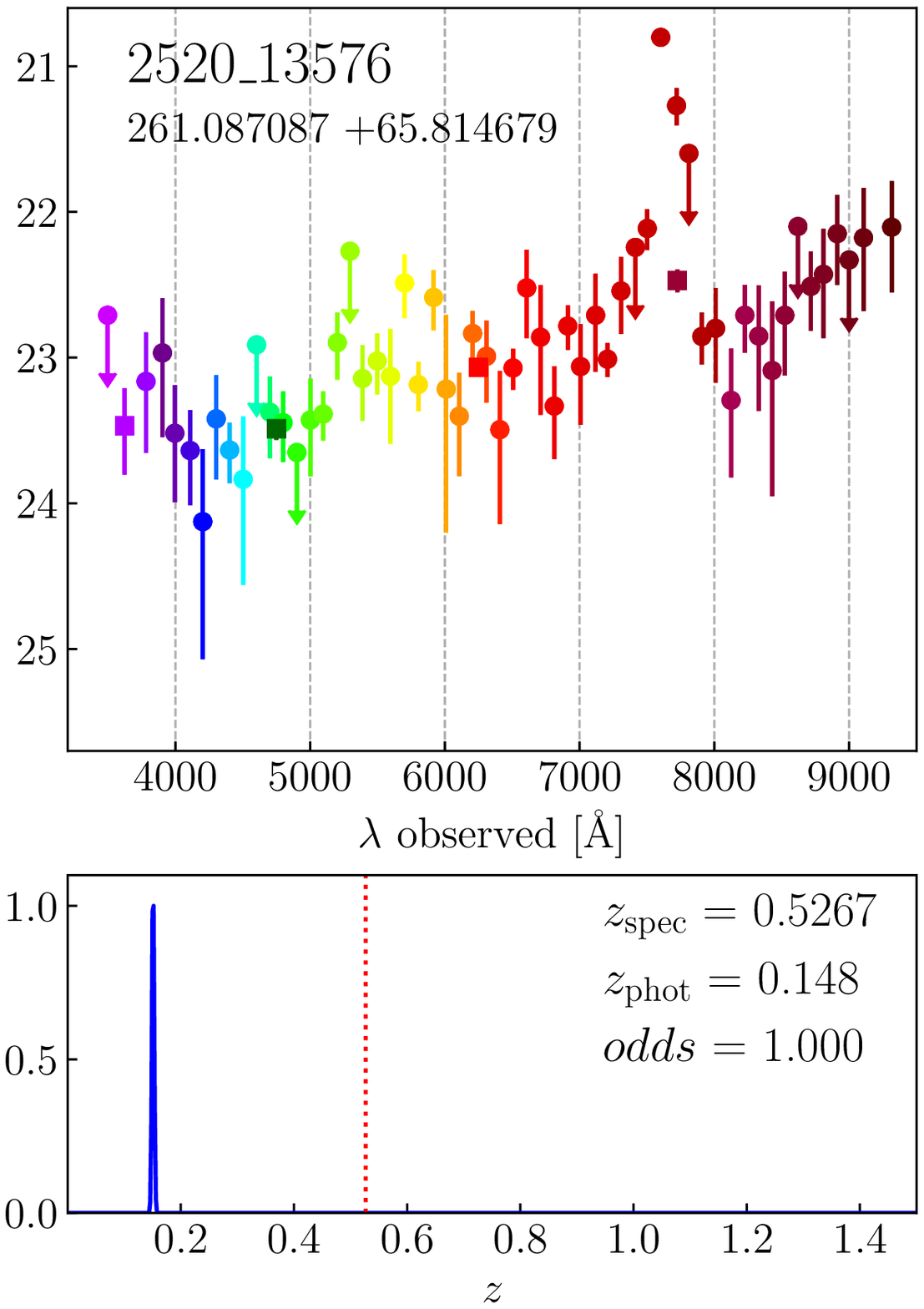}};
\draw (2.0,0.3) node[inner sep=0]{\includegraphics[height=1.7cm]{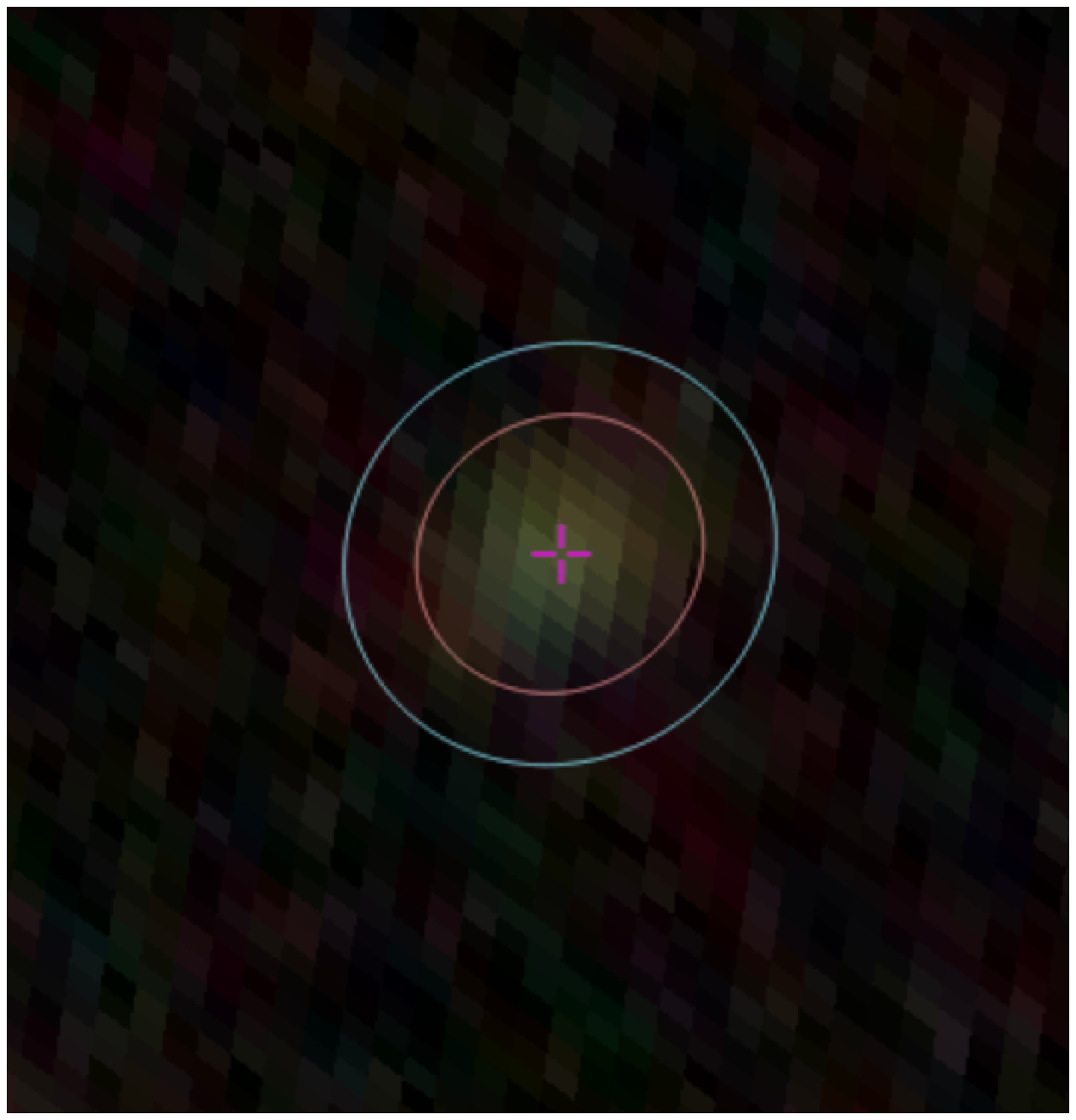}};
\end{tikzpicture}
\includegraphics[width=18.3cm]{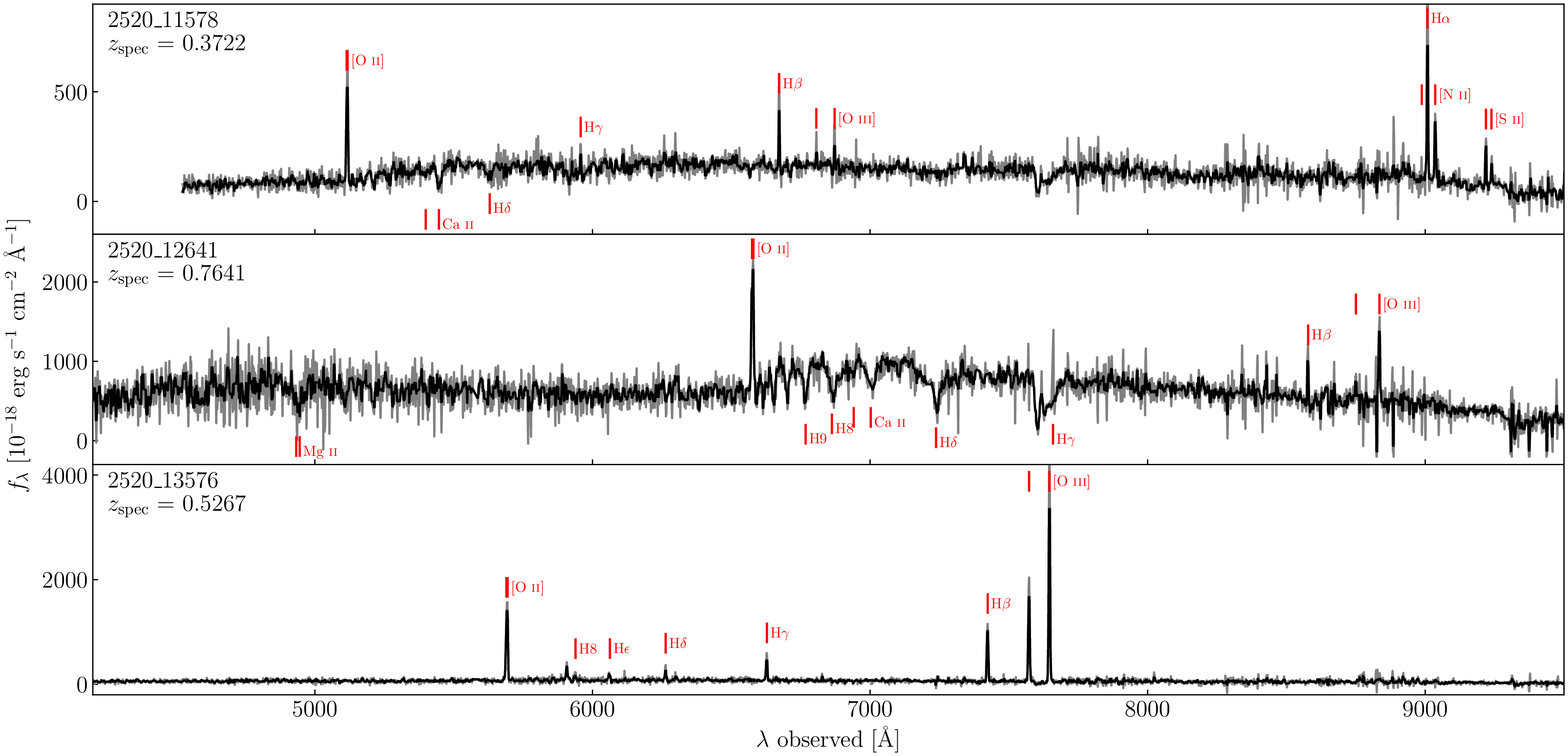}
\end{center}
\caption[]{Examples of J-NEP galaxies with high confidence photo-$z$ at an incorrect redshift. The top panels show their 60-band spectral energy distributions from J-PAS. Coloured circles (squares) indicate the PSFCOR magnitude in the narrow- (broad-) band filters. Error bars and arrows represent 1-$\sigma$ uncertainties and 3-$\sigma$ upper limits, respectively. Cutout images are a composite of the $g$ (blue), $r$ (green), and $i$ (red) bands. The red ellipses represent the 2 Kron radii apertures used to extract AUTO photometry. The PSFCOR apertures are half their size (1 Kron radius).
The middle panels show the normalised $z$PDF (solid line). The spectroscopic redshift is indicated with a vertical dotted line. The bottom panel shows the spectra from MMT/Binospec (Willmer et al. in preparation). The original spectrum (grey line) has been smoothed with a 5-element median filter to improve visualisation (black line). Red vertical lines mark some of the emission and absorption lines.\label{fig:outlier-examples}}
\end{figure*}

Figure \ref{fig:eta-odds-flags} shows that flagged sources have a $\sim$10\% chance for  catastrophic redshift errors additional to the outlier probability predicted from the $odds$. There is no clear dependence with the type of flags or the $odds$ value, except for sources near saturated stars (MASK\_FLAGS = 2), which at low $odds$ are almost invariably outliers.  
The $\sim$10\% increase in $\eta$ implies that the photo-$z$ of flagged sources with very high $odds$ are not as reliable as the $odds$ suggests. Therefore, flagged sources should be excluded from the sample selection for those applications that demand a very low outlier rate ($\eta$ $\lesssim$ 0.1). The top left panel in Fig. \ref{fig:outlier-examples} shows an example of a galaxy contaminated by a nearby star that obtains a very high confidence photo-$z$ at a wrong redshift.

\begin{figure} 
\begin{center}
\includegraphics[width=8.4cm]{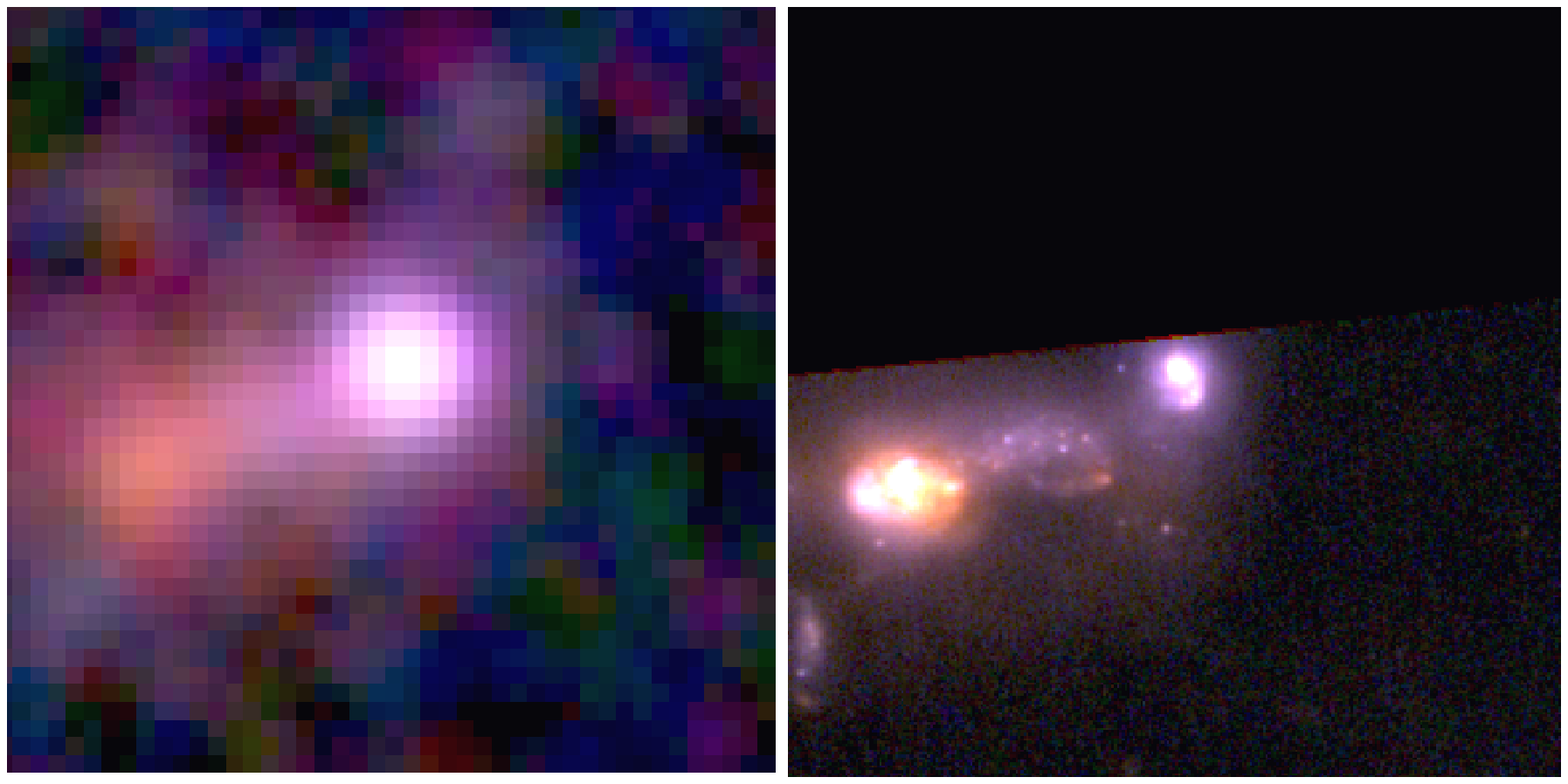}
\end{center}
\caption[]{10$\times$10\arcsec cutouts centred around the coordinates of the source 2520\_12641. The left panel is a composite of the J-NEP bands $g$ (blue), $r$ (green), and $i$ (red), smoothed with a top-hat filter. The right panel is a composite of the JWST/NIRCAM bands F090W (blue), F150W (green), and F200W (red) from JWST proposal
2738 \citep{Windhorst23}. North is up, East is left.\label{fig:JNEP-vs-NIRCAM}}
\end{figure}

The third cause for outliers is exotic spectra not represented in the template set.
By definition, those sources will have a poor fit at all redshifts. However, this does not prevent them from obtaining high $odds$, since the shape of the $z$PDF depends on the relative (not absolute) values of $\mathcal{L}$($z$) at each $z$.
In particular, both stars and quasars often obtain very high $odds$ (including $odds$ = 1) at wrong redshifts. However, we do not consider them as legitimate outliers since, by construction, our photo-$z$ measurements are conditional to the sources actually being galaxies.

In some cases, {\sc SExtractor} can detect close pairs of stars as a single extended source, which can get a very low $P_{star}$ value. A related case is that of a galaxy with a star in the foreground, or two galaxies at different redshifts that overlap in projection and get classified as a single extended source. We consider these as detection errors or contamination issues, not exotic spectra. 
An interesting example is shown in the middle panel of Fig. \ref{fig:outlier-examples}. {\sc SExtractor} detects as an elongated source what is actually two or three galaxies in close proximity, as shown in superb detail in the JWST/NIRCAM images (Fig. \ref{fig:JNEP-vs-NIRCAM}). The centroid of the J-NEP detection is very close to the bluer galaxy to the West, which might dominate the flux in the PSFCOR aperture. The Binospec spectrum, centred on the same galaxy, shows a post-starburst spectrum with strong Mg {\sc ii} $\lambda\lambda$2796, 2803 \AA{} and Balmer absorption lines as well as [O {\sc ii}] and [O {\sc iii}] emission. Other post-starburst galaxies that obtain their best fit with the same spectral template have accurate photo-$z$, suggesting that contamination from the red galaxy to the East is the cause for the outlier. We note that while the peak at $z$=0.832 completely dominates the $z$PDF, there is a second peak with $P$($z$)=0.025 that is consistent with the spectroscopic redshift.

A legitimate example of a galaxy with an exotic spectrum resulting in a wrong photo-$z$ at high $odds$ is shown in the top right panel of Fig. \ref{fig:outlier-examples}.   
The source is barely resolved in the J-NEP images ($P_{star}$=0.77) and is classified as a star by SDSS. However, the Binospec spectrum shows very strong [O {\sc iii}] $\lambda\lambda$4959, 5007 \AA{} lines (which cause the excess flux observed in the J0760 and J0770 bands). H$\alpha$ is redshifted out of the J-PAS and spectroscopic wavelength ranges, but the high [O {\sc iii}] equivalent width and the colours in the WISE bands suggest it is an AGN (albeit it is not detected by NuSTAR). 
Since there are no templates with such strong [O {\sc iii}] emission in our template set, {\sc LePhare} misidentifies the excess in J0760 and J0770 as H$\alpha$ at $z$$\sim$0.15.

\section{Summary and conclusions}\label{sec:summary}

This paper presents the photometry and photometric redshifts of J-NEP, a survey of $\sim$0.3 deg$^2$ centred on the JWST Time Domain Field in the North Ecliptic Pole performed with the Pathfinder camera on the JST/T250 telescope. 
Similarly to miniJPAS \citep{Bonoli21}, J-NEP observes in the 56 narrow-bands of J-PAS, plus $u$, $g$, $r$, and $i$, but reaches $\sim$0.5--1.0 magnitudes deeper, depending on the filter, with a typical PSF FWHM of $\sim$1.2\arcsec.

J-NEP provides single-band source catalogues in the 60 filters as well as forced (dual mode) photometry in several apertures for the 24,618 sources detected in the $r$-band image. All sources are classified as stars or galaxies using the neural network classifier of {\sc SExtractor} and the star probabilities of the Stellar-Galaxy Locus Classification of \citet{Lopez-Sanjuan19b}. 

We present a new method for correcting systematic offsets in the colours from PSF-corrected (PSFCOR) photometry, dubbed ``galaxy locus recalibration''. The method relies on the recalibration method proposed in \citetalias{HC21} to determine the median colours of galaxies (the ``galaxy locus'') in a reference sample. Then, it computes the magnitude offsets that, when applied to the observed photometry, make the median colours match the galaxy locus. 
While this new recalibration strategy is less accurate, it is applicable to any J-PAS pointing regardless of the availability of spectroscopy.  
 
We obtain photo-$z$ for all 6,662 J-NEP sources brighter than $r$=23. 
The photo-$z$ are computed by a customised version of {\sc LePhare} with the same configuration options used for miniJPAS, including the same 50 galaxy templates generated from a selection of miniJPAS galaxies and redshift priors derived from VVDS. 
The distribution of photo-$z$ for J-NEP galaxies is consistent with that found in miniJPAS, except for faint ($r$$>$22) sources, where J-NEP shows an excess of sources at $z_{\rm{phot}}$$<$0.2 and a deficit at $z_{\rm{phot}}$$>$1 relative to miniJPAS. Both are likely due to higher S/N in the photometry allowing for more robust photo-$z$ estimates.

We use a sample of 787 galaxies with reliable spectroscopic redshifts taken from the redshift catalogue in Willmer et al. (in preparation) to validate the accuracy of J-NEP photo-$z$. No spectroscopic information on the J-NEP sources was available to the photo-$z$ team before completion of the final photo-$z$ catalogue. This allows us to ensure that the photo-$z$ accuracy measured in the spectroscopic subsample is not affected by overfitting. 76\% of the spectroscopic sources have photo-$z$ error $\Delta z$$<$0.03, consistent with the expectation from the average $odds$ of the sample $\langle odds \rangle$ = 0.771.

The typical $\Delta z$ ($\sigma_{\rm{NMAD}}$) and the frequency of outliers ($\eta$) are slightly higher compared to miniJPAS for $r$$<$21.5, probably due to larger systematic errors, but they are up to $\sim$30--50\% lower at fainter magnitudes. This suggests that the higher S/N resulting from deeper observations, particularly in the redder bands, has a strong impact in the photo-$z$ accuracy of faint sources.

Despite the differences in depth and systematic uncertainties, the dependence of $\sigma_{\rm{NMAD}}$ with $odds$ in J-NEP is consistent within the uncertainties with that measured in miniJPAS. Furthermore, we predict the distribution of $\Delta z$ for the entire J-NEP sample using as reference the spectroscopic samples of J-NEP and miniJPAS and arrive at very similar results. 
Combined with a lack of dependence of $\sigma_{\rm{NMAD}}$ on the magnitude or the spectral type of the sources at constant $odds$, this suggests that we will be able to predict the $\sigma_{\rm{NMAD}}$ of any set of J-PAS sources from their $odds$ distribution alone, with no need for additional spectroscopy to calibrate the relation between $odds$ and $\Delta z$. 

We explore the factors contributing to photo-$z$ outliers in J-NEP and miniJPAS. For faint sources, the main reason is colour-space degeneracy due to large uncertainties in the observed colours. This is exacerbated in low $z$ dwarf galaxies by an unfavourable redshift prior and at high $z$ by the shift of key spectral features outside of the J-PAS wavelength range. At brighter magnitudes, artefacts in the photometry (including those signalled by the photometry flags) and exotic spectra not represented in the template set contribute most of the outliers with high $odds$.

The photometry and photo-$z$ presented in this paper are being used in multiple science projects by the J-PAS collaboration, including the search and characterisation of Lyman alpha emitters in miniJPAS and J-NEP (Torralba et al. in preparation), a census of extreme emission line galaxies at $z$$<$0.8 (Arroyo Polonio et al. in preparation), the characterisation of the stellar population and emission line properties of J-NEP galaxies (Gonz\'alez Delgado et al. in preparation) and the 3-D distribution and clustering properties of J-NEP galaxies (Maturi et al. in preparation).

All miniJPAS data, including images, catalogues, and value-added data products, are publicly available through the CEFCA catalogues portal\footnote{\url{http://archive.cefca.es/catalogues}}. 
Access to J-NEP data can be granted for specific use upon request to
the science directors of J-PAS prior to the official public release
currently scheduled for the third quarter of 2023. The JWST data are publicly available from the Space Telescope Science Institute MAST Archive\footnote{\url{https://doi.org/10.17909/d4w2-dz48}}.

\begin{acknowledgements}

We thank the anonymous referee for useful comments and suggestions that helped improve this work. We also thank R. Jansen for providing the outlines of the HST and JWST imaging used in Figure \ref{fig:footprint}. This paper has gone through internal review by the J-PAS collaboration.
Based on observations made with the JST/T250 telescope at the Observatorio Astrof\'isico de Javalambre (OAJ), in Teruel, owned, managed, and operated by the Centro de Estudios de F\'isica del Cosmos de Arag\'on (CEFCA). We acknowledge the OAJ Data Processing
and Archiving Unit (UPAD) for reducing and calibrating the OAJ data used in this work.
Funding for the J-PAS Project has been provided by the Governments of Spain and Arag\'on through the Fondo de Inversi\'on de Teruel, European FEDER funding and the Spanish Ministry of Science, Innovation and Universities, and by the Brazilian agencies FINEP, FAPESP, FAPERJ and by the National Observatory of Brazil. Additional funding was also provided by the Tartu Observatory and by the J-PAS Chinese Astronomical Consortium.
Funding for OAJ, UPAD, and CEFCA has been provided by the Governments of Spain and Arag\'on through the Fondo de Inversiones de Teruel; the Arag\'on Government through the Research Groups E96, E103, and E16\_17R; the Spanish Ministry of Science, Innovation and Universities (MCIU/AEI/FEDER, UE) with grant PGC2018-097585-B-C21; the Spanish Ministry of Economy and Competitiveness (MINECO/FEDER, UE) under AYA2015-66211-C2-1-P, AYA2015-66211-C2-2, AYA2012-30789, and ICTS-2009-14; and European FEDER funding (FCDD10-4E-867, FCDD13-4E-2685).
Partly based on observations taken at the MMT observatory, a joint facility operated by the Univesity of Arizona and the Smithsonian Institution. CNAW acknowledges support from NIRCam Development Contract NAS5-02105 from NASA Goddard Space Flight Center to the University of Arizona and from the HST-GO-15278.008 grant awarded by the Space Telescope Science Institute to the University of Arizona.
R.M.G.D. acknowledges financial support from the State Agency for Research of the Spanish MCIU through the "Center of Excellence Severo Ochoa" award to the Instituto de Astrofísica de Andalucía (SEV-2017-0709), and to  PID2019-109067-GB100.
Part of this work was supported by institutional research funding IUT40-2, JPUT907 and PRG1006 of the Estonian Ministry of Education and Research. We acknowledge the support by the Centre of Excellence “Dark side of the Universe” (TK133) financed by the European Union through the European Regional Development Fund.
L.S.J. acknowledges the support from CNPq (308994/2021-3) and FAPESP (2011/51680-6). A.F.-S. acknowledges support from project PID2019-109592GB-I00/AEI/10.13039/501100011033 (Spanish Ministerio de Ciencia e Innovaci\'on - Agencia Estatal de Investigaci\'on) and Generalitat Valenciana project of excellence Prometeo/2020/085. 
A.E. and J.A.F.O. acknowledge the financial support from the Spanish Ministry of Science and Innovation and the European Union NextGenerationEU through the Recovery and Resilience Facility project ICTS-MRR-2021-03-CEFCA.
This study forms part of the Astrophysics and High Energy Physics programme and was supported by MCIN with funding from European Union NextGenerationEU (PRTR-C17.I1) and by Generalitat Valenciana under the project n. ASFAE/2022/025.

\end{acknowledgements}

\appendix 

\section{Accuracy of recalibration with the galaxy locus}\label{sec:appendix}

\begin{figure} 
\begin{center}
\includegraphics[width=8.4cm]{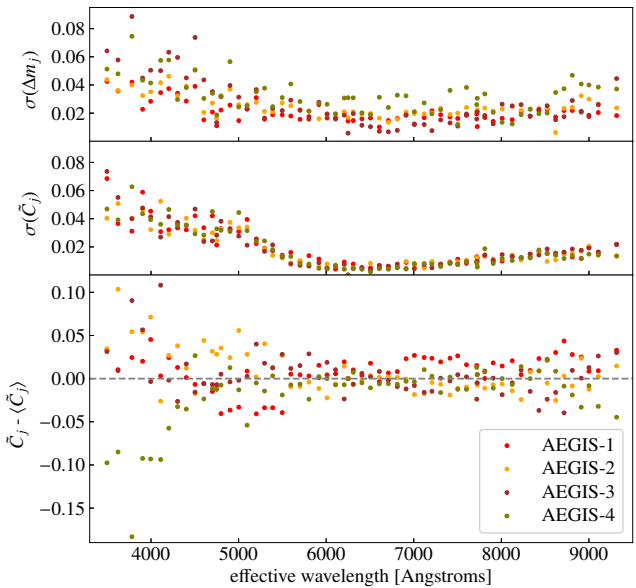}
\end{center}
\caption[]{Top panel: 1-$\sigma$ uncertainty in the recalibration offsets for individual miniJPAS pointings (from \citetalias{HC21}). Middle panel: 1-$\sigma$ uncertainty in the median observed colours, $C_j$, of the galaxies with $r$$<$21 in each miniJPAS pointing. Bottom panel: dispersion between the galaxy loci obtained on individual miniJPAS pointings.\label{fig:gl-residuals}}
\end{figure} 

In the ISMFR method based on SPS model fits that was presented in \citetalias{HC21}, the magnitude offset for band $j$ in a given pointing, $\Delta m_j$, is computed as the median difference between the observed magnitudes and those predicted from the SPS models for a sample of spectroscopic galaxies. The uncertainty in $\Delta m_j$, $\sigma$($\Delta m_j$), depends on the photometric errors of the individual galaxies and the number of spectroscopic galaxies available (note that only $r$$<$22 galaxies are used due to very large photometric errors for the narrow bands in the fainter ones).
The statistical uncertainty in the recalibration with this method is $\sigma$($\Delta m_j$) $\sim$ 0.02 mag except for the bluest bands, where the photometric errors of individual galaxies are larger (see top panel in Fig. \ref{fig:gl-residuals}). 

The galaxy locus recalibration (GLR) inherits this uncertainty as a systematic offset in the colour of all the galaxies in a given pointing. In addition, GLR is affected by the statistical uncertainty, $\sigma$($\tilde{C}_j$), in the determination of the median colour of the $r$$<$21 galaxy sample. We estimate $\sigma$($\tilde{C}_j$) for each miniJPAS pointing with bootstrap resampling (middle panel). The total uncertainty of the galaxy locus defined on a single pointing is thus

\begin{equation}
\sigma_{tot}(\tilde{C}_j) = \sqrt{\sigma^2(\Delta m_j) + \sigma^2(\tilde{C}_j)}
\end{equation}

Using the galaxy locus defined from one pointing to recalibrate another one means that cosmic variance may also have an impact on the results.
The bottom panel in Fig. \ref{fig:gl-residuals} shows the dispersion in $\tilde{C}_j$ values among the individual miniJPAS pointings. The median absolute difference in $\tilde{C}_j$ between any two miniJPAS pointings is 0.031 mag (standard deviation: 0.046 mag), again dominated by the bluer bands.
To test whether cosmic variance or other sources of error not included in $\sigma_{tot}$($\tilde{C}_j$) contribute significantly to this dispersion, we compute for each pair ($a$, $b$) of miniJPAS pointings the reduced $\chi^2$ statistic

\begin{equation}
\chi^2_r(a,b) = \frac{1}{n_j-1}\sum_{j=1}^{n_j} \frac{(\tilde{C}_j^a-\tilde{C}_j^b)^2}{\sigma^2_{tot}(\tilde{C}_j^a) + \sigma^2_{tot}(\tilde{C}_j^b)}
\end{equation}

The values of $\chi^2_r$ range from 0.54 to 1.04, suggesting that differences in $\tilde{C}_j$ between pointings are entirely accounted for by the uncertainties of the ISMFR method and the statistical uncertainty in the median colours.

\begin{figure} 
\begin{center}
\includegraphics[width=8.4cm]{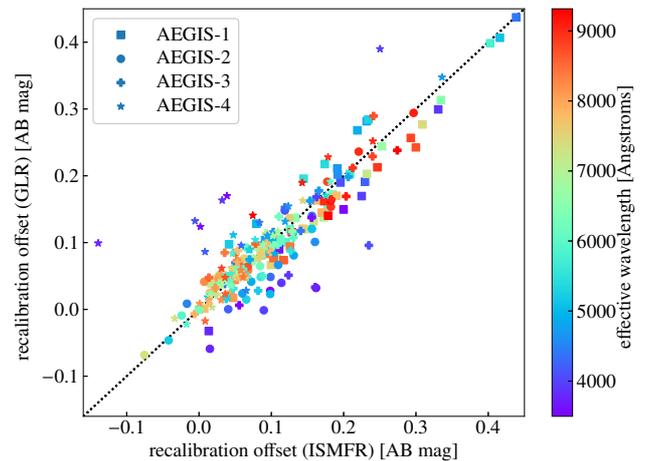}
\end{center}
\caption[]{Comparison of the magnitude corrections determined for each band for individual miniJPAS pointings using the original recalibration method presented in \citetalias{HC21} (ISMFR) and the recalibration with galaxy locus from this work (GLR).\label{fig:compare-recalibrations}}
\end{figure}

As a final test, we compute recalibration offsets with the GLR method for all four miniJPAS pointings. For each of them we use the galaxy locus obtained from the other three pointings. In Fig. \ref{fig:compare-recalibrations} we compare the magnitude offsets obtained with the ISMFR and GLR methods. The 1-$\sigma$ dispersion in the difference between the two methods is 0.039 mag, dominated by the bluest bands. If we consider only bands with $\lambda_{eff}$$>$4500 \AA{} the dispersion decreases to 0.024 mag. This is comparable to the uncertainty in the ISMFR method and implies that the total systematic error in galaxy colours with the GLR method is at most $\sim$0.04--0.05 mag. 

\begin{thebibliography}{999}
\bibitem[Arnouts, \& Ilbert(2011)]{Arnouts11}Arnouts, S., \& Ilbert, O.\ 2011, LePHARE: Photometric Analysis for Redshift Estimate, ascl:1108.009
\bibitem[Baqui et al.(2021)]{Baqui21}Baqui, P.~O., Marra, V., Casarini, L. et al. 2021, A\&A, 645, 87B
\bibitem[Ben{\'i}tez et al.(2009)]{Benitez09}Ben\'itez, N., Moles, M., Aguerri, J. A. L., et al. 2009, ApJ, 692, L5
\bibitem[Ben{\'i}tez et al.(2014)]{Benitez14} Ben\'itez, N., Dupke, R., Moles, M., et al. 2014, arXiv e-prints, arXiv:1403.5237
\bibitem[Bertin \& Arnouts(1996)]{Bertin96}Bertin, E. \& Arnouts, S. 1996, A\&AS, 117, 393
\bibitem[Bertin et al.(2002)]{Bertin02} Bertin, E., Mellier, Y., Radovich, M., et al.\ 2002, Astronomical Data Analysis Software and Systems XI, 281, 228
\bibitem[Bertin(2006)]{Bertin06} Bertin, E.\ 2006, Astronomical Data Analysis Software and Systems XV, 351, 112
\bibitem[Bonoli et al.(2021)]{Bonoli21} Bonoli, S., Mar{\'\i}n-Franch, A., Varela, J., et al.\ 2021, A\&A, 653, A31
\bibitem[Boquien et al.(2019)]{Boquien19} Boquien, M., Burgarella, D., Roehlly, Y., et al.\ 2019, A\&A, 622, A103
\bibitem[Cenarro et al.(2014)]{Cenarro14} Cenarro, A.~J., Moles, M., Mar{\'\i}n-Franch, A., et al.\ 2014, SPIE Conf. Ser., 9149, 91491I
\bibitem[Chaves-Montero et al.(2022)]{Chaves-Montero22} Chaves-Montero, J., Bonoli, S., Trakhtenbrot, B., et al.\ 2022, A\&A, 660, A95
\bibitem[Coe et al.(2006)]{Coe06} Coe, D., Ben{\'\i}tez, N., S{\'a}nchez, S.~F., et al.\ 2006, AJ, 132, 926
\bibitem[Fabricant et al.(2019)]{Fabricant19} Fabricant, D., Fata, R., Epps, H., et al.\ 2019, PASP, 131, 075004
\bibitem[Gonz{\'a}lez Delgado et al.(2021)]{Gonzalez-Delgado21} Gonz{\'a}lez Delgado, R.~M., D{\'\i}az-Garc{\'\i}a, L.~A., de Amorim, A., et al.\ 2021, A\&A, 649, A79
\bibitem[Gonz{\'a}lez Delgado et al.(2022)]{Gonzalez-Delgado22} Gonz{\'a}lez Delgado, R.~M., Rodr{\'\i}guez-Mart{\'\i}n, J.~E., D{\'\i}az-Garc{\'\i}a, L.~A., et al.\ 2022, A\&A, 666, A84
\bibitem[Green et al.(2018)]{Green18}Green, G. M., Schlafly, E. F., Finkbeiner, D., et al. 2018, MNRAS, 478, 651
\bibitem[Hern\'an-Caballero et al.(2021)]{HC21}Hern\'an-Caballero, A., et al.\ 2021, A\&A, 654, 101
\bibitem[Iglesias-P{\'a}ramo et al.(2022)]{Iglesias-Paramo22}Iglesias-P{\'a}ramo, J., Arroyo, A., Kehrig, C., et al. 2022, A\&A, 
\bibitem[Jansen \& Windhorst(2018)]{Jansen18} Jansen, R.~A. \& Windhorst, R.~A.\ 2018, PASP, 130, 124001
\bibitem[Kansky et al.(2019)]{Kansky19} Kansky, J., Chilingarian, I., Fabricant, D., et al.\ 2019, PASP, 131, 075005
\bibitem[Laur et al.(2022)]{Laur22} Laur, J., Tempel, E., Tamm, A., et al.\ 2022, A\&A, 668, A8
\bibitem[Le F{\`e}vre et al.(2005)]{LeFevre05} Le F{\`e}vre, O., Vettolani, G., Garilli, B., et al.\ 2005, A\&A, 439, 845
\bibitem[L{\'o}pez-Sanjuan et al.(2019a)]{Lopez-Sanjuan19a}L\'opez-Sanjuan, C., Varela, J., Crist\'obal-Hornillos, D., et al. 2019a, A\&A, 631, A119
\bibitem[L{\'o}pez-Sanjuan et al.(2019b)]{Lopez-Sanjuan19b} L{\'o}pez-Sanjuan, C., V{\'a}zquez Rami{\'o}, H., Varela, J., et al.\ 2019b, A\&A, 622, A177 
\bibitem[L{\'o}pez-Sanjuan et al.(2021)]{Lopez-Sanjuan21} L{\'o}pez-Sanjuan, C., Yuan, H., V{\'a}zquez Rami{\'o}, H., et al.\ 2021, A\&A, 654, A61
\bibitem[Mart{\'\i}nez-Solaeche et al.(2021)]{Martinez-Solaeche21} Mart{\'\i}nez-Solaeche, G., Gonz{\'a}lez Delgado, R.~M., Garc{\'\i}a-Benito, R., et al.\ 2021, A\&A, 647, A158
\bibitem[Mart{\'\i}nez-Solaeche et al.(2022)]{Martinez-Solaeche22} Mart{\'\i}nez-Solaeche, G., Gonz{\'a}lez Delgado, R.~M., Garc{\'\i}a-Benito, R., et al.\ 2022a, A\&A, 661, A99
\bibitem[Molino et al.(2019)]{Molino19}Molino, A., Costa-Duarte, M. V., Mendes de Oliveira, C., et al. 2019, A\&A, 622, A178
\bibitem[Newman et al.(2013)]{Newman13} Newman, J.~A., Cooper, M.~C., Davis, M., et al.\ 2013, ApJS, 208, 5
\bibitem[Polsterer et al.(2016)]{Polsterer16} Polsterer, K.~L., D'Isanto, A., \& Gieseke, F.\ 2016, arXiv:1608.08016
\bibitem[Queiroz et al.(2022)]{Queiroz22}Queiroz, C., Abramo, L.~R., Rodrigues, N.~V.~N., et al., 2022, MNRAS in press (arXiv:2202.00103)
\bibitem[Rodr{\'\i}guez-Mart{\'\i}n et al.(2022)]{Rodriguez-Martin22} Rodr{\'\i}guez-Mart{\'\i}n, J.~E., Gonz{\'a}lez Delgado, R.~M., Mart{\'\i}nez-Solaeche, G., et al.\ 2022, A\&A, 666, A160
\bibitem[Schlafly et al.(2016)]{Schlafly16}Schlafly, E. F., Meisner, A. M., Stutz, A. M., et al. 2016, ApJ, 821, 78
\bibitem[Whitten et al.(2019)]{Whitten19}Whitten, D. D., Placco, V. M., Beers, T. C., et al. 2019, A\&A, 622, A182
\bibitem[Windhorst et al.(2023)]{Windhorst23} Windhorst, R.~A., Cohen, S.~H., Jansen, R.~A., et al.\ 2023, AJ, 165, 13
\bibitem[Xiao et al.(2021)]{Xiao21} Xiao, K., Yuan, H., Varela, J., et al.\ 2021, ApJS, 257, 31

\end{thebibliography}
\end{document}